\documentclass[10pt,journal]{IEEEtran}
\ifCLASSINFOpdf
\usepackage[pdftex]{graphicx}
\else
\usepackage[dvips]{graphicx}
\fi
\usepackage{amsmath}
\usepackage{algorithmic}

\usepackage{array}

\ifCLASSOPTIONcompsoc
\usepackage[caption=false,font=normalsize,labelfont=sf,textfont=sf]{subfig}
\else
\usepackage[caption=false,font=footnotesize]{subfig}
\fi

\usepackage{stfloats}
\usepackage{url}

\usepackage{algorithm}

\usepackage[numbers, compress]{natbib}

\usepackage[utf8]{inputenc} 
\usepackage[T1]{fontenc}    
\usepackage[pdftex, pdftitle={Towards Robust Waveform-Based Acoustic Models}, pdfauthor={Dino Oglic}, colorlinks=true, bookmarksnumbered, citecolor=darkblue, urlcolor=darkblue, linkcolor=darkblue]{hyperref}       

\usepackage{booktabs}       
\usepackage{amsfonts}       
\usepackage{nicefrac}       
\usepackage{microtype}      

\usepackage{amssymb}
\usepackage{amsthm}
\usepackage{thmtools,thm-restate}
\usepackage{mathtools}

\usepackage{makeidx}

\usepackage{calc}
\usepackage{multirow}
\usepackage{makecell}

\usepackage{wrapfig}
\usepackage[font=footnotesize, skip=3pt]{caption}

\usepackage{color}
\usepackage{xcolor}
\usepackage{colortbl}
\definecolor{lb}{RGB}{155,206,227}
\definecolor{db}{RGB}{31,120,180}
\definecolor{lg}{RGB}{178,223,138}
\definecolor{dg}{RGB}{51,160,44}
\definecolor{darkblue}{rgb}{0,0,0.75}

\definecolor{caddendum}{RGB}{0,0,0}
\definecolor{cerratum}{RGB}{0,0,0}
\definecolor{c-lss-1}{RGB}{28,144,153}
\definecolor{c-lss}{RGB}{255,0,0}
\definecolor{c-dpp}{RGB}{127,121,73}
\definecolor{c-dpp-1}{RGB}{127,121,73}
\definecolor{c-km}{RGB}{117,107,177}
\definecolor{lloyd-km}{RGB}{241,163,64}

\newtheorem{definition}{Definition}
\newtheorem{theorem}{Theorem}
\newtheorem*{theorem*}{Theorem}

\DeclareMathOperator*{\argmax}{arg\,max}

\DeclareMathOperator*{\tr}{tr}

\newcommand{\D}[1]{\,\mathrm{d}{#1}}
\newcommand{\brackets}[1]{\left( {#1} \right)}
\newcommand{\Brackets}[1]{\Big( {#1} \Big)}
\newcommand{\cbrackets}[1]{\left\{ {#1} \right\} }
\newcommand{\sbrackets}[1]{\left[ {#1} \right]}

\newcommand{\norm}[1]{\left\Vert {#1} \right\Vert}
\newcommand{\absolute}[1]{\left\vert {#1} \right\vert}

\newcommand{\ceq}[1]{(\ref{#1})}

\newlength\figureheight
\newlength\figurewidth

\usepackage{tikz}
\usepackage{pgfplots}
\usepackage{pgf}
\usepackage{adjustbox}
\usepackage{pgfplotstable}
\pgfplotsset{compat=newest}
\pgfplotsset{
	tick label style={font=\small},
	label style={font=\small},
	legend style={font=\small},
	every axis/.append style={
		thick,
		tick style={semithick, black},
		axis line style={-},
		axis x line =bottom,
		axis y line =left
	}
}

\usepgfplotslibrary{groupplots}
\usepgfplotslibrary{fillbetween}
\usepgfplotslibrary{external} 
\usepackage{soul}

\begin{document}
%
\title{Towards Robust Waveform-Based Acoustic Models}
%
%
%
	
\author{Dino~Oglic,
	Zoran~Cvetkovic,
	Peter~Sollich, Steve~Renals,~and Bin~Yu
	\thanks{This work was supported in part by EPSRC under Grant EP/R012067/1.}
	\thanks{D.\ Oglic is with the Applied Analytics and AI, Data Sciences and AI, BioPharmaceuticals R\&D, AstraZeneca, Cambridge CB2 8PA, UK.  This work was done in part while he was with the Department of Engineering, King's College London, London WC2R 2LS, UK. Correspondence to: $\texttt{dino.oglic@astrazeneca.com}$.}
	\thanks{Z.\ Cvetkovic is with the Department of Engineering, King's College London, London WC2R 2LS, UK (e-mail: zoran.cvetkovic@kcl.ac.uk).}%
	\thanks{P.\ Sollich is with the Department of Mathematics, King's College London, London WC2R 2LS, UK, and also the Institute for Theoretical Physics, University of G\"ottingen, 37073 G\"ottingen, Germany (e-mail: peter.sollich@kcl.ac.uk).}%
	\thanks{S.\ Renals is with the Center for Speech Technology Research, University of Edinburgh, Edinburgh EH8 9AB, UK (e-mail: s.renals@ed.ac.uk).}
	\thanks{B. Yu is with the Departments of Statistics and Electrical Engineering and Computer Sciences, UC Berkeley, Berkeley, CA 94720, USA (e-mail: binyu@stat.berkeley.edu).}
}
	
%
%

\markboth{IEEE Transactions on Audio, Speech, and Language Processing}%
{Shell \MakeLowercase{\textit{et al.}}: Bare Demo of IEEEtran.cls for IEEE Journals}
%

\IEEEpubid{\copyright~2022 IEEE. Citation information: \href{https://doi.org/10.1109/TASLP.2022.3172632}{DOI 10.1109/TASLP.2022.3172632}, IEEE/ACM Transactions on Audio, Speech, and Language Processing.}


\maketitle
\IEEEpeerreviewmaketitle

\begin{abstract}
We study the problem of learning robust acoustic models in adverse environments, characterized by a significant mismatch between training and test conditions. This problem is of paramount importance for the deployment of speech recognition systems that need to perform well in unseen environments. 
First, we characterize data augmentation theoretically as an instance of vicinal risk minimization, which aims at improving risk estimates during training by replacing the delta functions that define the empirical density over the input space
with an approximation of the marginal population density in the vicinity of the training samples. More specifically, we assume that local neighborhoods centered at training samples can be approximated using a mixture of Gaussians, and demonstrate theoretically that this can incorporate robust inductive bias into the learning process. We then specify the individual mixture components implicitly via data augmentation schemes, designed to address common sources of spurious correlations in acoustic models. To avoid potential confounding effects on robustness due to information loss, which has been associated with standard feature extraction techniques (e.g., \textsc{fbank} and \textsc{mfcc} features), we focus 
on the waveform-based setting. Our empirical results show that the approach can generalize to unseen noise conditions, with $150\%$ relative improvement in out-of-distribution generalization compared to training using the standard risk minimization principle. Moreover, the results demonstrate competitive performance relative to models learned using a training sample designed to match the acoustic conditions characteristic of test utterances. 
\end{abstract}
	

%
\IEEEpeerreviewmaketitle

\section{Introduction}
\label{sec:intro}

We consider the problem of improving the performance of 
acoustic models in adverse environments, where there is a significant mismatch between training and testing acoustic conditions. This problem is of paramount importance for the deployment of automatic speech recognition systems that are required to perform well in unseen environments, without suffering any significant performance degradation. Recently, there has been considerable progress in improving the performance of acoustic models in adverse conditions for filterbank features~\cite[e.g., see][]{yu2013feature,Vincent17,specaugment,banpassaug}. However, there are still significant performance gaps for testing environments characterized as novel relative to the ones seen during training. To the best of our knowledge, there has not been a study on the impact of adverse environments on waveform-based acoustic models~\cite{sincnet,edinburgh,parznets20}, typically encountered in truly end-to-end speech recognition systems. 
The waveform-based setting is particularly interesting because it allows for an empirical evaluation free of confounding effects on robustness due to information loss and non-adaptive feature extraction process.
More specifically, several comparative studies of automatic and human speech recognition~\cite{alsteris,meyer07,petere99} suggest that the information loss inherent to filterbank features can adversely affect robustness to standard environmental distortions arising from additive and channel (linear filtering) noise~\cite{Ager11,Yousafzai11a}. 
The main challenge comes from the high variability of speech signals that are representative of a sub-phonetic unit. This can be caused by differences between speakers (e.g., accents, pronunciations, speaking styles, emotional states, etc.), environmental noise, different microphones, reverberation, and recording devices~\cite{yu2013feature}. Moreover, the sources of variability are typically non-stationary and can interact with speech signals in a non-linear way~\cite{Yoshioka12}. Hence, it is difficult if not impossible to avoid a mismatch between training and testing environments.\IEEEpubidadjcol

At the core of modern automatic speech recognition systems are deep learning models that exploit associations between frame representations and corresponding sub-phonetic units when learning to generalize from training data to the population level. A problem arises when such associations are characteristic of the training sample but are not present in the test samples or, in general, at the population level. This phenomenon is known as spurious correlation and it hinders the generalization abilities of acoustic models. Typically, spurious correlations are scrambled by augmenting the training samples with acoustic conditions that resemble those expected within testing environments. In this regard, a particularly influential learning regime is multi-condition/style training where inputs are transformed by naturally occurring additive noise signals with various signal-to-noise ratios~\cite{Vincent17,banpassaug,Cui15}. The noise types are usually selected such that they reduce the mismatch between training and testing conditions; there are several publicly available databases of naturally occurring environment noise signals. This type of data augmentation is practical because it preserves sub-phonetic labels assigned to frames of the original speech signal (i.e., there is no need for re-alignment of augmented utterances). While multi-condition training allows for significant performance improvement in approximately matched testing conditions, the  
error rates are severely increased in environments with novel speech variability characteristics. Thus, the main shortcoming of such approaches is the fact that it is infeasible to enumerate all possible noise types combined with different signal-to-noise ratios, as well as other sources of variability that one can expect in testing environments~\cite{Li14}. 

To tackle this issue and allow for learning of robust acoustic models,  
we propose an approach based on vicinal risk minimization, which aims at improving risk estimates during training by replacing the delta functions that define the empirical density over the input space with an approximation of the population density in the vicinity of the training samples. We assume that the vicinal densities can be described using a mixture of Gaussians and characterize the individual mixture components implicitly via data augmentation schemes that are designed to scramble frequent sources of spurious correlations in speech (Section~\ref{sec:data-augmentation}). In our approach, we rely on white noise as a source of randomness and couple it with  
linear transformations of the input signal that are based on band-pass and band-stop filtering. This purely synthetic  approach differs from prior work that typically resorts to databases of standard environment noise types and/or acoustic conditions. The focus of our study is on two types of mismatch between training and test environments. In the first case, we are interested in improving out-of-distribution generalization of waveform-based models when the difference between the two environments lies in the level and type of background noise. The second case, on the other hand, deals with spurious correlations introduced by different microphones used for recording. The latter is challenging to address for waveform-based models due to the fact that the feature extraction process is fully automated and utterance level mean-normalization cannot be performed as in the case of standard non-adaptive filterbank features. Prior work has established that such normalizations can be fundamental in dealing with spurious correlations introduced by different microphones and stationary signal corruptions~\cite{Vincent17,neethu20}.

Our theoretical contributions include a characterization of robustness relative to the waveform domain, given in terms of the Jacobian and Hessian tensors of the sufficient statistic (Section~\ref{sec:theory}). In addition to this, we provide insights into how vicinal risk minimization affects the learning process in terms of inductive bias --- data fitting at the level of local neighborhoods centered at training observations instead of pointwise mappings characteristic of standard empirical risk minimization. 

The results of our empirical evaluation are presented in Section~\ref{sec:exps}. The main focus of the evaluation is on the generalization from clean to noisy speech, with a severe mismatch between training and run-time conditions.  
To avoid possible confounding effects~\cite{alsteris,meyer07,petere99} of the standard feature extraction process on the robustness 
our empirical study quantifies the effectiveness of the approach relative to the waveform-based models.  
We conduct our analysis on \textsc{aurora4} using the Kaldi \emph{clean-condition recipe} and a recently proposed neural architecture for waveform-based speech recognition called \textsc{parznets 2d}~\cite{parznets20}. Our empirical results demonstrate that the proposed approach can improve out-of-distribution generalization abilities of models trained on clean speech by more than $150\%$ in relative terms. Moreover, the results obtained are competitive with the best possible augmentation principle where training samples are transformed using noise types that appear in the test fold.
Having demonstrated that the approach can aid generalization to unseen conditions, we evaluate it on conversational speech, thus showing its utility for generalization from headset recorded to distant-talking speech.

\begin{figure*}[t]
	\centering
	\includegraphics[scale=0.2]{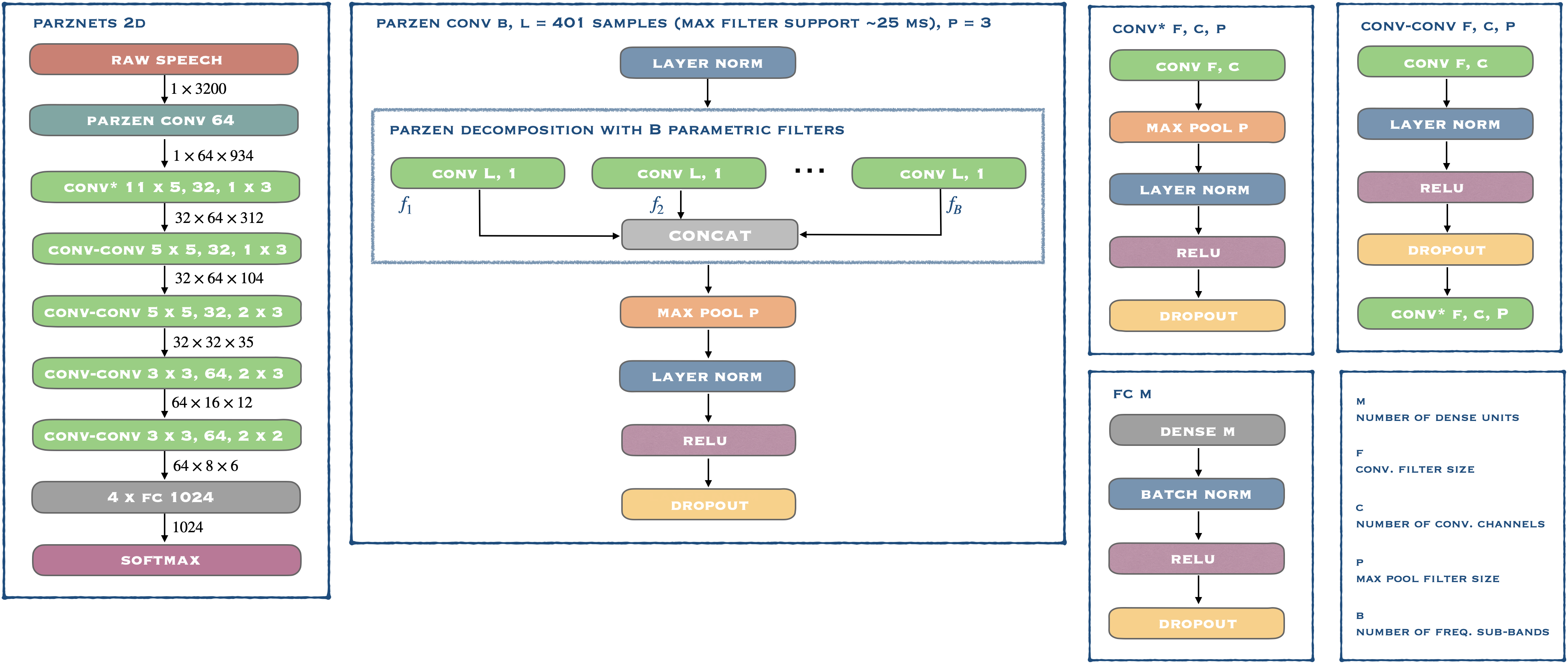}
	\caption{The figure describes the architecture for \textsc{parznets} with \textsc{2d} convolutional operators. This is supplemented with an illustration of the Parzen convolutional block that decomposes a raw speech frame into frequency sub-bands.}
	\label{fig:arch}\vspace{-2ex}
\end{figure*}

\section{Parzen Filters and Parznets}
\label{sec:parznets}

In this section, we provide a brief review of Parzen filters and \textsc{parznets 2d}~\cite{parznets20} -- the waveform based neural architecture used in our empirical evaluations. The proposed data augmentation schemes depend in part on band-pass filtering and for simplicity we have used Parzen filters as a realization of this operator. \textsc{parznets 2d} is a neural architecture that provides an effective means for fully automated feature extraction directly from speech signals. This is achieved by the first layer of the network, which is defined using parametric convolutions that allow for efficient learning of band-pass filters.

In speech recognition, band-pass filtering of signals is traditionally performed by weighted averaging of power spectra,  
computed over segments of fixed duration. Alternatively, the signal can be convolved with a filter directly in the time-domain. Motivated by this, the first layer of \textsc{parznets 2d} is designed to emulate this operation via a parametric time-domain convolutional operator. To that end, the network employs a family of differentiable band-pass filters based on cosine modulations of compactly supported Parzen windows~\cite{parzen1962}. In particular, our empirical analysis employs a squared Epanechnikov window function~\cite{epanechnikov1969nonparametric}
\begin{align*}
	k_{\gamma}\brackets{t} = \left\{ \begin{array}{ll}
		\brackets{1 - \gamma t^2}^2 & \absolute{t} \leq \nicefrac{1}{\sqrt{\gamma}} \\
		0 & \text{otherwise} \ , \\ 
	\end{array}   \right.
\end{align*}
where $\gamma$ is a parameter controlling the window width. 
To allow for flexible placement of the center/modulation frequency, the filterbank relies on 
cosine modulation. Thus, Parzen filters are defined with only two differentiable parameters, $\eta$ controlling the modulation frequency and $\gamma$ controlling the filter bandwidth,
\begin{align}
	\label{eq:parzen}
	\phi_{\eta,\gamma}\brackets{t}=\cos \brackets{2\pi \eta t} \cdot k_{\gamma}\brackets{t} \ . 
\end{align}
As illustrated in~Figure~\ref{fig:arch}, for each filter configuration $\cbrackets{\brackets{\eta_i, \gamma_i}}_{i=1}^B$, Eq.~\ceq{eq:parzen} is used to generate a one dimensional filter with maximum length given by the number of samples in $25$ ms of speech; filters with shorter support are symmetrically padded with zeros. 

\textsc{parznets 2d} take the outputs of parametric convolutions and concatenate them into a spectro-temporal decomposition of a signal, which is then passed to a max pooling operator, followed by layer normalization~\cite{layernorm}.  
The outputs of the parametric convolutional block  
are then passed to a sequence of standard convolutional operators that perform further band-pass filtering and compression of the signal by different max pooling operators. The convolutional blocks generate a set of \emph{automatically extracted features}, which are then passed to a multi-layer perceptron with four hidden layers.

\section{Theory}
\label{sec:theory}

Let $\mathcal{X} \subset \mathbb{R}^d$ be a compact set containing raw speech frames of fixed duration in its interior (e.g., $200$ ms long frames of speech) and $\mathcal{Y}$ the space of categorical labels (e.g., state ids in hybrid \textsc{hmm}-\textsc{dnn} models). Suppose that a set of labeled examples $\cbrackets{\brackets{x_i, y_i}}_{i=1}^n$ has been drawn independently from a latent Borel probability measure defined on $\mathcal{X} \times \mathcal{Y}$.

Deep learning models with \textsc{softmax} outputs (e.g., feedforward neural architectures) typically assume that the conditional probability of a label $y \in \mathcal{Y}$ given an instance $x \in \mathcal{X}$ can be approximated with an exponential family model~\citep{jaynes57} 
\begin{align}
p\brackets{y \mid x, \alpha, W} = \frac{\exp \brackets{\alpha_y^{\top} \Psi\brackets{x \mid W}}}{\sum_{y' \in \mathcal{Y}} \exp \brackets{\alpha_{y'}^{\top} \Psi\brackets{x \mid W}}} \ ,
\label{Pygivenx}
\end{align}
where $\alpha \in \mathbb{R}^{D \times \absolute{\mathcal{Y}}}$ is a parameter matrix with columns $ \alpha_y \in \mathbb{R}^D$ defining the \emph{softmax probabilities} $p\brackets{y \mid x,\alpha, W}$ and $\Psi\brackets{x \mid W} \in \mathbb{R}^{D}$ is a sufficient statistic of $x$, given by pre-softmax neural network parameters $W$.
This model can also be written as a special case of a conditional exponential family model~\cite{altun04}
\begin{align*}
p\brackets{y \mid x, \alpha, W}=\frac{\exp \brackets{\alpha^{\top} \Phi\brackets{x,y \mid W}}\ }{\ \sum_{y' \in \mathcal{Y}} \exp \brackets{\alpha^{\top} \Phi\brackets{x,y' \mid W}}} \ ,
\end{align*}
where $\alpha \in \mathbb{R}^{D \cdot \absolute{\mathcal{Y}}}$ now denotes a parameter vector defining the \emph{softmax probabilities} and $\Phi\brackets{x,y \mid W} \in \mathbb{R}^{D \cdot \absolute{\mathcal{Y}}}$ is a sufficient statistic of $y \mid x$.  
If the latter is selected such that $\Phi\brackets{x,y \mid W}=\mathrm{vec}(\mathbf{e}_y\ \Psi\brackets{x \mid W}^{\top})$, where $\mathbf{e}_y \in \mathbb{R}^{\absolute{\mathcal{Y}}}$ is the \emph{one-hot} column vector having one at the position of a categorical label $y \in \mathcal{Y}$ and zero elsewhere, then the simpler form (\ref{Pygivenx}) is retrieved. Here, $\mathrm{vec}\brackets{\cdot}$ denotes the operator that concatenates rows of a matrix into a vector.  
In more complex deep learning architectures such as \emph{sequence-to-sequence} and \emph{attention} models~\citep[e.g., see][]{transformers}, one typically designs a model-specific sufficient statistic $\Phi\brackets{x, y \mid W}$ using the decoder component. The latter takes as input the corresponding categorical label $y \in \mathcal{Y}$ along with a hidden state representation of the input sequence produced by an encoder.

\subsection{Vicinal Risk Minimization}
\label{subsec:vicinal-risk-minimization}

In empirical risk minimization, the learning algorithm selects a hypothesis that minimizes a loss/risk function with respect to the empirical distribution given by delta functions located at the training samples. In the case of acoustic models, one typically minimizes the negative log-likelihood, i.e.,
\begin{align*}
\mathcal{R}_{emp}\brackets{W, \alpha} = -\frac{1}{n}\sum_{i=1}^n \log p\brackets{y_i \mid x_i, \alpha, W} \ .
\end{align*}
As the ultimate goal of a learning algorithm is to select a hypothesis that minimizes the expected risk, the vicinal risk minimization~\cite{vicinal} aims at improved estimates by replacing the delta functions with some approximation of the density in the vicinity of training instances. More formally, the vicinal variant of the negative log-likelihood is given by
\begin{align*}
\mathcal{R}_{vic}\brackets{W, \alpha} =- \frac{1}{n}\sum_{i=1}^n \int \log p\brackets{y_i \mid x, \alpha, W} \D{P_{x_i}\brackets{x}} \ ,
\end{align*}
where $P_{x_i}\brackets{\cdot}$ is a density estimate in the vicinity of $x_i$. 

Data augmentation can be seen as an instance of vicinal risk minimization, where local density estimates are designed to improve the generalization properties of the empirical estimators. To illustrate this, we assume that the density estimate in the vicinity of $x_i$ can be approximated by a mixture of Gaussians, i.e.,
\begin{align}
\label{eq:gauss-mixture}
P_{x_i}\brackets{x}=\frac{1}{K}\sum_{k=1}^K \mathcal{N}\brackets{\mu_{ik}, \sigma_k^2 \mathbb{I}} \ ,
\end{align}
with $\mu_{ik} \coloneqq \mu_{ik}\brackets{x_i} \in \mathbb{R}^d$ and $\sigma_k \in \mathbb{R}$. In Section~\ref{sec:data-augmentation}, we propose several data augmentation schemes and formally introduce these mean functions as linear transformations of training samples. The variance parameter is included here to control the signal-to-noise ratio and, thus, the level of robustness associated with each of the mixture components.

The vicinal risk induced by $P_{x_i}$ is then given by 
\begin{align*}
\ell_n \brackets{W, \alpha} = -\frac{1}{nK} \sum_{i,k} \mathbb{E}_{x \sim \mathcal{N}\brackets{\mu_{ik}, \sigma_k^2 \mathbb{I}}} \sbrackets{\log p\brackets{y_i \mid x, \alpha, W}}  \ ,
\end{align*}
with $1\leq i \leq n$ and $1\leq k \leq K$. As the integral defining the expectation operator is intractable, we resort to Monte Carlo approximation with a small number of \textsc{iid} samples 
\begin{align}
\label{eq:vicinal-empirical}
\ell_{n,m} \brackets{W, \alpha} = -\frac{1}{mn} \sum_{i=1}^n \sum_{j=1}^m \log p\brackets{y_i \mid x_{ij}, \alpha, W} \ ,
\end{align}
where $x_{ij} = \mu_{ik'} + \epsilon_{ik'}$ is a sample from the mixture model that corresponds to instance $x_i$, realized by first selecting uniformly a mixture component $k' \sim \mathcal{U}_{\cbrackets{1,\dots, K}}$ and then the offset $\epsilon_{ik'} \sim \mathcal{N}\brackets{0, \sigma_{k'}^2\mathbb{I}}$.
Due to the fact that $\log$ is a concave function, we have from the Jensen inequality that 
\begin{align*}
\ell_n \brackets{W, \alpha} \geq -\frac{1}{n} \sum_{i=1}^n \log \mathbb{E}_{x \sim P_{x_i}\brackets{x}} \sbrackets{p\brackets{y_i \mid x, \alpha, W}} \ .
\end{align*}
Hence, the vicinal risk function is an upper bound on the negative log-expected likelihoods over neighborhoods centered at training instances and minimization of such an objective is likely to promote locally smooth hypotheses.

For density estimates $P_{x_i}$ defined with an isotropic zero-mean Gaussian distribution,~Chapelle et al.~\cite[Section 3,][]{vicinal} have demonstrated that the notion of vicinal risk amounts to introducing a penalty term given by the squared norm of the parameter vector, i.e., the standard $\ell_2$-regularization mechanism for linear models. Thus, learning a neural network by optimizing the vicinal risk from Eq.~\ceq{eq:vicinal-empirical} can be seen as an extension of the standard regularized risk minimization principle that works effectively for linear models.

An alternative insight into the properties incorporated into the learning algorithms via vicinal risk minimization can be obtained by introducing a temperature parameter $T$ into the likelihood function: 
\begin{align*}
p_T\brackets{y \mid x, \alpha, W} = \frac{\exp \brackets{\alpha_y^{\top} \Psi\brackets{x \mid W}/ T}}{\sum_{y' \in \mathcal{Y}} \exp \brackets{\alpha_{y'}^{\top} \Psi\brackets{x \mid W}/ T} } \ .
\end{align*}
Then, we have that 
$\lim_{T \rightarrow 0}\ p_T\brackets{y \mid x, \alpha, W} = \xi_y$
with 
\begin{align*}
\xi_y = \left\{
\begin{array}{ll}
1  & \text{if} \ y = \argmax_{y' \in \mathcal{Y}} \ p \brackets{y' \mid x, \alpha, W} \\
0 & \text{otherwise} \ .
\end{array}
\right.
\end{align*}
From the latter expression it follows that for each mixture component from Eq.~\ceq{eq:gauss-mixture}
\begin{align*}
\begin{aligned}
& \mathbb{E}_{\epsilon \sim \mathcal{N}\brackets{0, \sigma^2\mathbb{I}} } \sbrackets{\lim_{T \rightarrow 0} \ p_T\brackets{y \mid x + \epsilon, \alpha, W}} & \\
& = P_\epsilon \brackets{y = \argmax_{y' \in \mathcal{Y}} p\brackets{y' \mid x + \epsilon, \alpha, W} } &
\end{aligned}
\end{align*}
In other words, minimization of the objective in Eq.~\ceq{eq:vicinal-empirical} can be seen as an approximation to the following problem
\begin{align*}
\min_{W, \alpha}\ - \sum_{i,j} \log P_{\epsilon_{ij}} \brackets{y_i = \argmax_{y' \in \mathcal{Y}} p\brackets{y' \mid \mu_{ij} + \epsilon_{ij}, \alpha, W} } ,
\end{align*}
which maximizes the likelihood of correct classification over neighborhoods rather than particular training instances. The latter estimator is known as \emph{randomized smoothing} in machine learning and several recent works have provided certifiable robustness bounds for that estimator~\cite[e.g., see][]{cohen19,Salman19}.

\subsection{Inductive Bias via Data Augmentation}
\label{subsec:inductive-bias}

In this section, we provide two further insights into the effects of data augmentation on the robustness of acoustic models. The technical derivations build on prior work by~\cite{dao19b} that focused on data augmentation for kernel machines. We extend those observations/results to neural networks and show from another perspective that learning with augmented samples promotes vicinal smoothness of the learned models/hypotheses.

We first recall that $\alpha \in \mathbb{R}^{D \times \absolute{Y}}$ is the \emph{softmax parameter matrix} with columns given by $\alpha_y \in \mathbb{R}^D$. The log-likelihood of label $y \in \mathcal{Y}$ conditioned on an instance $x \in \mathcal{X}$ can now be expressed using the following function
\begin{align*}
\begin{aligned}
& \tau (\zeta\brackets{x}, y) \coloneqq \alpha_y^{\top} \Psi\brackets{x \mid W} - \log \sum_{y' \in \mathcal{Y}} \exp \brackets{\alpha_{y'}^{\top} \Psi\brackets{x \mid W}} & \\
& = \zeta_y\brackets{x} - \log \sum_{y' \in \mathcal{Y}} \exp \Brackets{\zeta_{y'}\brackets{x}} \ , &
\end{aligned}
\end{align*}
where $\zeta\brackets{x} \coloneqq \alpha^{\top} \Psi\brackets{x \mid W} \in \mathbb{R}^{\absolute{\mathcal{Y}}}$ is a vector with components $\zeta_y\brackets{x} \in \mathbb{R}$. 
Now, recall also that the density in the vicinity of an instance is approximated using a mixture of Gaussians (see Eq.~\ref{eq:gauss-mixture}), 
which allows for a re-parametrization via $P\brackets{\epsilon} \coloneqq P_{x_i}\brackets{x_i + \epsilon}$ for $\epsilon \in \mathbb{R}^d$. 

Taking the first-order Taylor approximation of the function $\tau$ around $\zeta^{(0)}$ and passing it through the expectation operator with respect to the vicinal density $P\brackets{\epsilon}$, we obtain
\begin{align*}
\begin{aligned}
& \mathbb{E}_{\epsilon} \sbrackets{\tau \Brackets{\zeta\brackets{x + \epsilon}, y}} = & \\& \tau \brackets{\zeta^{(0)}, y} 
+ \nabla \tau\brackets{\zeta^{(0)}, y}^{\top} \mathbb{E}_{\epsilon} \sbrackets{\zeta\brackets{x+\epsilon} - \zeta^{(0)}} + & \\
&  o\brackets{\mathbb{E}_{\epsilon}\sbrackets{\norm{\zeta\brackets{x+\epsilon} - \zeta^{(0)}}}} \ .&
\end{aligned}
\end{align*}
Setting $\zeta^{(0)} = \mathbb{E}_{\epsilon}\sbrackets{\zeta\brackets{x+\epsilon}}$ we have 
 $   \mathbb{E}_{\epsilon}\sbrackets{\zeta\brackets{x+\epsilon} - \zeta^{(0)}}=0$, which
implies that the vicinal negative log-likelihood is approximately equal to
\begin{align*}
\begin{aligned}
& \ell_{n}\brackets{W, \alpha} \approx 
- \frac{1}{n} \sum_{i=1}^n \log p \Brackets{y_i \mid \mathbb{E}_{\epsilon}\sbrackets{\Psi\brackets{x_i+\epsilon \mid W}}, \alpha} \ . &
\end{aligned}
\end{align*}
Thus, the vicinal risk objective approximately maximizes the likelihood of the mean embedding of a training instance with respect to the density estimate $P_{x_i}$, which is also used as a data augmentation mechanism.

On the other hand, the second order Taylor approximation with respect to $\zeta^{(0)}$ gives
\begin{align*}
\begin{aligned}
& \mathbb{E}_{\epsilon} \sbrackets{\tau \Brackets{\zeta\brackets{x + \epsilon}, y}}  & \\
& \approx \tau \brackets{\zeta^{(0)}, y} + \frac{1}{2} \mathbb{E}_{\epsilon} \sbrackets{\Delta_{\zeta}\brackets{\epsilon}^{\top} \nabla^2 \tau \brackets{\zeta^{(0)}, y} \Delta_{\zeta}\brackets{\epsilon}}  & \\
& \leq \tau \brackets{ \zeta^{(0)}, y} + \frac{\mu}{2} \tr \brackets{ \mathbb{E}_{\epsilon}\sbrackets{\Delta_{\zeta}\brackets{\epsilon} \Delta_{\zeta}\brackets{\epsilon}^{\top}}} ,&
\end{aligned}
\end{align*}
where $\mu$ is the largest eigenvalue of Hessian matrix $\nabla^2 \tau\brackets{\zeta^{(0)}, y}$ and $\Delta_{\zeta}\brackets{\epsilon}=\zeta\brackets{x+\epsilon}-\zeta^{(0)}$. The second term is just the variance of the individual entries of the vector $\zeta\brackets{x+\epsilon} \in \mathbb{R}^{\absolute{\mathcal{Y}}}$, relative to the data augmentation distribution from Eq.~\ceq{eq:gauss-mixture}. The latter follows from the fact that $\Delta_{\zeta}\brackets{\epsilon}$ is a zero-mean random variable. 
Now, the vicinal negative log-likelihood is approximately equal to
\begin{align*}
\begin{aligned}
& \ell_{n} \brackets{W, \alpha} \approx - \frac{1}{n} \sum_{i=1}^n \log p \brackets{y_i \mid \mathbb{E}_{\epsilon} \sbrackets{\Psi \brackets{x_i + \epsilon \mid W}}, \alpha } & \\
& \qquad \qquad \quad + \mathbb{E}_{\epsilon} \sbrackets{ \norm{\Delta_{\zeta}\brackets{\epsilon}}^2_{\nabla^2 \tau \brackets{\zeta^{(0)}, y_i}}} \leq & \\
& - \frac{1}{n} \sum_{i=1}^n \log p \brackets{y_i \mid \mathbb{E}_{\epsilon} \sbrackets{\Psi \brackets{x_i + \epsilon \mid W}}, \alpha } + \frac{\mu}{2} \mathrm{Var}\sbrackets{\Delta_{\zeta}\brackets{\epsilon}} \ . &
\end{aligned}
\end{align*}
From here it follows that vicinal risk minimization optimizes a lower bound on the objective consisting of the negative log-likelihood and a variance-based penalty term. The latter is defined over the space of pre-softmax vectors and relative to the random variable defined by vicinal density. Thus, the second order approximation tells us that vicinal risk minimization aims at assigning a similar conditional distribution of labels given an instance, across a neighborhood specified by the data augmentation principle. 
As a result, the predictions are likely to remain the same in local neighborhoods/vicinity around training points which fosters locally robust sufficient statistic.

\subsection{A Notion of Locally Robust Sufficient Statistic}
\label{subsec:local-robustness}

In this section, we introduce a notion of a locally robust sufficient statistic $\Psi\brackets{x \mid W}$ and provide a bound on the deviation between its values over a neighborhood in the vicinity of a training sample. In our analysis, we focus on robustness relative to a ball of constant radius centered at a training sample, which contains the high density vicinal region described by Eq.~\ceq{eq:gauss-mixture} in its interior.
Henceforth, we will simplify our notation and denote this sufficient statistic with $\Psi\brackets{x}$ (omitting $W$). 

A robust representation of waveform signals $\Psi\brackets{x}$ should be stable with respect to additive noise perturbations. This can be, for instance, achieved with a contraction mapping. An operator $\Psi$ is said to be a contraction if there exists a positive constant $L < 1$ such that for all $x \in \mathcal{X}$ and $\epsilon \in \mathbb{R}^d$ with $x + \epsilon \in \mathcal{X}$
\begin{align*}
\norm{\Psi\brackets{x + \epsilon} - \Psi \brackets{x}} \leq L \norm{\epsilon} \ .
\end{align*}

The contraction property 
is a stability notion that holds across the whole space and represents a \emph{strong notion of robustness}. Typically, this is relaxed by requiring that the sufficient statistic is stable under small additive perturbations of the signal. More formally, for a constant $r>0$ and all $z \in \mathcal{B}\brackets{x, r}=\{x + \epsilon \in \mathcal{X} \mid \norm{\epsilon} < r\}$ it is required that
$\norm{\Psi\brackets{z} - \Psi \brackets{x}} < r$.

A more flexible way to express the latter notion of robustness is to assume that $\epsilon$ is a random variable that follows an isotropic Gaussian distribution, i.e., $\epsilon \sim \mathcal{N}\brackets{\epsilon \mid 0, \sigma^2\mathbb{I}}$, and require that for all $0<\delta < 1$ and some $r \coloneqq r\brackets{\sigma, \delta} > 0$ 
\begin{align*}
\mathbb{E}_{\epsilon}\sbrackets{\norm{\Psi\brackets{x + \epsilon} - \Psi \brackets{x}} < r} > 1 - \delta \ .
\end{align*}
This is equivalent to requiring that for all $0 < \delta < 1$ there exists $r > 0$ such that
$
P_{\epsilon}\brackets{\norm{\Psi\brackets{x + \epsilon} - \Psi \brackets{x}} \geq r} < \delta  .
$

Before proceeding with the bound on the deviation between values of a sufficient statistic over a neighborhood centered at a training sample, we introduce the relevant operators and notions that will be used to express this type of robustness.

\begin{definition}
\label{def:trace-constants}
Let $\nabla \Psi\brackets{x} \in \mathbb{R}^{d \times D}$ be the Jacobian matrix of the sufficient statistic $\Psi\brackets{x} \in \mathbb{R}^D$ at a training sample $x \in \mathbb{R}^d$, which is given by $\nabla\Psi_{ij}\brackets{x} = \frac{\partial}{\partial x_i}\Psi_j\brackets{x}$. Let also $\nabla^2\Psi\brackets{x} \in \mathbb{R}^{d \times d \times D}$ be the Hessian tensor of the same sufficient statistic given by $\nabla^2\Psi_{ik,j}\brackets{x}=\frac{\partial}{\partial x_i}\frac{\partial}{\partial x_k} \Psi_j\brackets{x}$ with $1 \leq i,k \leq d$ and $1\leq j\leq D$. The constants 
\begin{align*}
\begin{aligned}
& a \coloneqq \tr \brackets{\nabla \Psi\brackets{x}^{\top} \nabla \Psi \brackets{x}} \quad \text{and} & \\
& b \coloneqq \sum_{j=1}^D \tr \brackets{\nabla^2 \Psi_{**,j}\brackets{x}} + \tr\brackets{\nabla^2 \Psi_{**,j}\brackets{x}\nabla^2 \Psi_{**,j}\brackets{x}}  & 
\end{aligned}
\end{align*}
with $\Psi_{**,j}\brackets{x} \in \mathbb{R}^{d \times d}$, describe the spectra of Jacobian and Hessian tensors, thus encapsulating the geometry of the neighborhood around an instance $x \in \mathcal{X}$.
\end{definition}

The following theorem characterizes a class of sufficient statistics that satisfy the notion of local robustness introduced above (a proof of this result is given in the appendix).

\begin{theorem}
\label{thm: local-robust}
	Suppose that the Hessian tensor of sufficient statistic $\Psi\brackets{x}$ is uniformly bounded for all $x \in \mathcal{X}$. 
	Then, for all $\delta>0$  
	\begin{align*}
	P_{\epsilon}\sbrackets{\norm{\Psi\brackets{x + \epsilon} - \Psi \brackets{x}} < \frac{\sigma}{\delta} \brackets{\sqrt{a} + \sigma \sqrt{\nicefrac{b}{2}}}} > 1 - \delta \ .
	\end{align*}
\end{theorem}

A practical meaning of this result is that the pre-softmax layer outputs that correspond to speech signals sampled from the vicinal density around a training sample concentrate in the feature space given by that layer.

\section{WaveAugment: Data Augmentation for Learning Robust Acoustic Models}
\label{sec:data-augmentation}

\begin{figure}[t]
	\centering
	\includegraphics[scale=0.23]{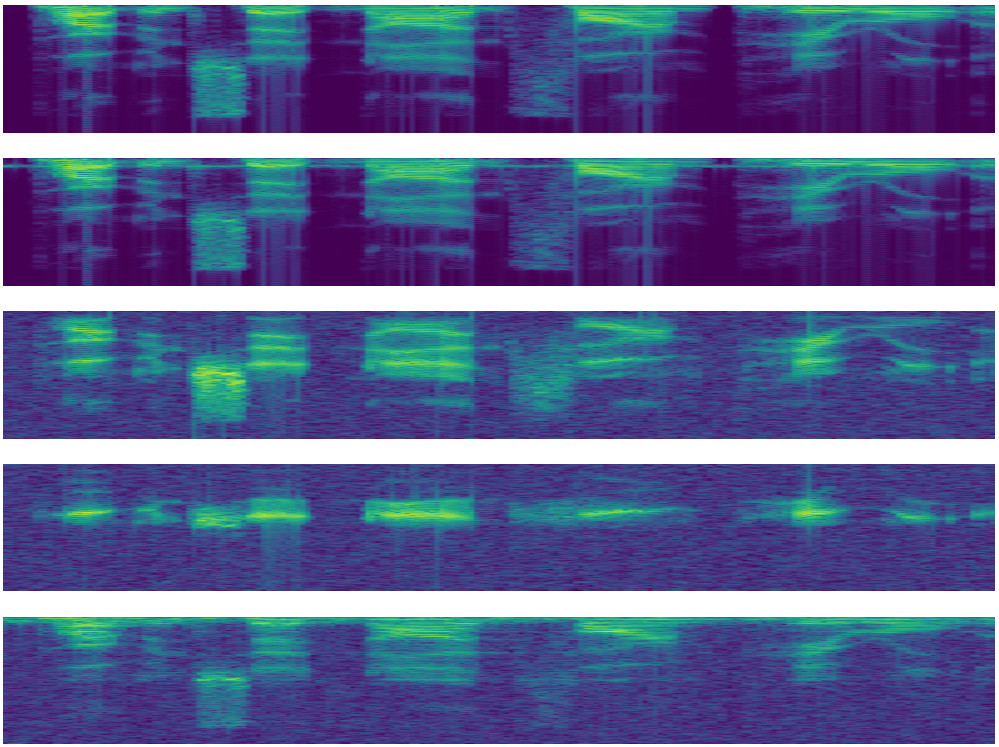}
	\caption{The figure illustrates perturbations to the magnitude spectrum of an utterance generated by the proposed algorithms. The panels (ordered from top to bottom) depict a clean speech utterance along with its perturbations by Algorithm~\ref{alg:bandlimited}-\ref{alg:noisy-rir}, respectively.}
	\label{fig:augmentations}\vspace{-2ex}
\end{figure}

In this section, we introduce four data augmentation schemes operating directly in the waveform domain with the goal of scrambling spurious correlations from the training sample and allowing for learning of robust acoustic models capable of generalizing to unseen environments. The main focus of our study is the improvement in out-of-distribution generalization when there is a mismatch between training and testing conditions that can be characterized by differences in background noise types, differences between training and run-time microphones, and  room reverberation. In the remainder of the section, we provide algorithmic descriptions and motivation for the proposed data augmentation schemes based on insights from signal processing and the observed effectiveness in our empirical study (see Section~\ref{sec:exps} for details). Figure~\ref{fig:augmentations} illustrates the effects of the proposed data augmentation schemes on the magnitude spectrum of a clean speech utterance from \textsc{aurora4}~\cite{aurora4}.

\subsection{Band-limited White Noise}
\label{subsec:bandlimited}

We start with an augmentation scheme that transforms an input speech waveform by adding a noise signal with support over low-frequency components only. The main motivation for this is to scramble spurious correlations that may be present in the training sample and can impede generalization to certain noise environments that affect this part of the spectrum (e.g., babble, airport, or car noise). The underlying source of randomness is white noise and we use different signal-to-noise ratios to further diversify the resulting signal corruptions.

\begin{algorithm}[t]
	\algsetup{linenosize=\tiny}
	\caption{\textsc{band-limited white noise}} 
	\label{alg:bandlimited}
	\begin{algorithmic}[1]
		{
			\fontsize{9}{11}\selectfont
			\REQUIRE 
			audio signal $x\brackets{t}$, sampling rate $f$, bandpass frequency range given by $\omega_{\min}$ and $\omega_{\max}$, filterbank size $p$, filter support size $T$, \textsc{snr} range given by $\gamma_{\min}$ and $\gamma_{\max}$\vspace*{1ex}  
			\STATE $\{\omega_i\}_{i=1}^p \leftarrow \textsc{evenly\_spaced\_modes} \brackets{\omega_{\min}, \omega_{\max}, f, p}$\vspace*{0.5ex}
			\STATE $\{\xi_i\}_{i=1}^p \leftarrow \textsc{bandwidths} \brackets{\{\omega_i \}_{i=1}^p, \omega_{\min}, \omega_{\max}, f}$\vspace*{0.5ex}
			\STATE $(\omega, \xi) \sim \mathcal{U}_{\cbrackets{(\omega_1, \xi_1), \dots, (\omega_p, \xi_p)}}$\vspace*{0.5ex}
			\STATE $h\brackets{t} \leftarrow \textsc{parzen\_filter}\brackets{\omega, \xi, T}$\vspace*{0.5ex}
			\STATE $\epsilon\brackets{t} \leftarrow \brackets{\epsilon_0 * h}\brackets{t}$ with $\epsilon_0(t) \sim \mathcal{N}(0, 1)$\vspace*{0.5ex}
			\STATE $\epsilon\brackets{t} \leftarrow \textsc{snr\_scale} \brackets{\epsilon\brackets{t}, \gamma}$  with $\gamma \sim \mathcal{U}\brackets{\gamma_{\min},\gamma_{\max}}$\vspace*{0.5ex}
			\STATE \textbf{return} $x\brackets{t}+\epsilon\brackets{t}$
		}
	\end{algorithmic}
\end{algorithm}

Algorithm~\ref{alg:bandlimited} provides a pseudo-code description of this augmentation scheme. In steps $1$ and $2$, we set $p$ evenly spaced modulation frequencies $\omega_i$ and corresponding bandwidths $\xi_i$ of  Parzen filters which are defined via cosine modulations of the squared Epanechnikov window (see Section~\ref{sec:parznets}). These are bandpass filters with support over the low-frequency part of the spectrum, with the lowest and highest frequencies specified as input to the algorithm. Following this, step $3$ selects one such filter uniformly at random. The selected filter is then convolved with a white noise signal (step $5$) so that the resulting additive noise vector has support over low-frequency components alone. As the final step in the generation of band-limited white noise, the algorithm selects a signal-to-noise ratio  uniformly at random from a given input range (step $6$). The output of the algorithm is an additive corruption of the input speech waveform signal. In our implementation, we have used $p=8$ Parzen filters and restricted the modulation range by setting $\omega_{\min}=50$ and $\omega_{\max}=800$ Hz. The corresponding bandwidths $\xi_i$ are set to be equal and jointly cover the specified frequency band, 
that is, $\xi_i \approx (\omega_{\max}-\omega_{\min})/p$.
For the signal-to-noise ratio range, we have opted for corruptions in the range of $8$-$32$ dB.

We conclude with a reference to our theoretical considerations from Section~\ref{sec:theory}. This scheme does not involve any change to the input signal prior to adding the additive noise term. As a result, it corresponds to a Gaussian mixture component with the mean parameter given by the training sample (i.e., input signal to Algorithm~\ref{alg:bandlimited}) itself and a non-isotropic variance term (due to the convolution with the low-pass filter) that accounts for the appropriate signal-to-noise ratio (see also Eq.~\ref{eq:gauss-mixture}).

\subsection{Notch Filtered Signals}
\label{subsec:notch}

In this section, we propose a data augmentation scheme based on notch filters~\cite{wang2013} that removes certain frequencies from the input and replaces them with white noise. The main motivation behind this scheme is to bias the learning process away from certain types of spurious correlations by embedding a noise signal, which is independent of sub-phonetic labels, into frequency ranges susceptible to environment effects. There are two use cases that we would like to target with this type of augmentation. The first one deals with low-frequency interference due to microphones and the second one with scrambling of the high frequency content of the signal that is susceptible to environment effects (e.g., street or car noise).

A notch filter is defined by a frequency that sets a dip in the spectrum around which the content is eliminated. 
We use three-tap notch filters of the form
\begin{align}
\label{eq:notch-filter}
h_{\omega} \sbrackets{t} = \left\{ \begin{array}{ll}
1 & t=\pm1\ ,\\
-2\cos \brackets{\omega} & t=0\ ,\\
0 & \text{otherwise} \ , \\ 
\end{array}   \right.
\end{align} 
where $\omega$ is the location of the \emph{frequency dip}.
 In our particular case, we would like to introduce two frequency dips, one at zero to tackle the microphone effects and another in the high frequency range (e.g., above $5$ kHz).

\begin{algorithm}[t]
	\algsetup{linenosize=\tiny}
	\caption{\textsc{noisy double-dip cosine notch}} 
	\label{alg:double-notch}
	\begin{algorithmic}[1]
		{
			\fontsize{9}{11}\selectfont
			\REQUIRE 
			audio signal $x\brackets{t}$, sampling rate $f$, notch frequency range given by $\omega_{\min}$ and $\omega_{\max}$, number of high frequency notch filters $p$, \textsc{snr} range given by $\gamma_{\min}$ and $\gamma_{\max}$\vspace*{1ex}  
			\STATE $\{\omega_i\}_{i=1}^p \leftarrow \textsc{evenly\_spaced\_modes} \brackets{\omega_{\min}, \omega_{\max}, f, p}$\vspace*{0.5ex}
			\STATE $\omega \sim \mathcal{U}_{\{\omega_1, \dots, \omega_p \}}$ with high freq. notch filter $h_{\omega}$ (see Eq.~\ref{eq:notch-filter}) \vspace*{0.5ex}
			\STATE $z\brackets{t} \leftarrow \brackets{x * h_0}\brackets{t}$ with notch filter $h_{0}$ (see Eq.~\ref{eq:notch-filter})\vspace*{0.5ex}
			\STATE $z\brackets{t} \leftarrow \brackets{z * h_{\omega}}\brackets{t}$ and $\epsilon(t) \sim \mathcal{N}(0, 1)$\vspace*{0.5ex}
			\STATE $\epsilon\brackets{t} \leftarrow \textsc{snr\_scale} \brackets{\epsilon\brackets{t}, \gamma}$  with $\gamma \sim \mathcal{U}\brackets{\gamma_{\min},\gamma_{\max}}$\vspace*{0.5ex}
			\STATE \textbf{return} $z\brackets{t}+\epsilon\brackets{t}$
		}
	\end{algorithmic}
\end{algorithm}

Algorithm~\ref{alg:double-notch} provides a pseudo-code description of this signal transformation. It takes as input the sampling rate along with high frequency notch range given by $\omega_{\min}$ and $\omega_{\max}$, the number of notch frequencies $p$, and signal-to-noise ratio range that will dictate the magnitude of white noise added to the filtered signal. In the first step, the algorithm creates a set of frequencies in the specified range from which a high frequency notch is selected uniformly at random (step $2$). 
Following this, the notch filter with a dip at zero is convolved with the original signal (step $3$). The algorithm then takes the selected high frequency dip (see step $2$) and convolves the resulting (zero notched) signal with the second notch filter defined by it (step $4$). As a result, the signal now has frequency dips at zero and the selected high frequency. In the next step, we generate a white noise and select a signal-to-noise ratio uniformly at random from the range provided as input to the algorithm (step $5$). The output of the algorithm is an additive corruption of the notch filtered signal. As the white noise has support over the whole frequency range, this scheme injects noise signal that is independent of sub-phonetic labels at the selected \emph{frequency dips} of the original signal and in this way fosters robustness.
In our implementation, we restrict the high frequency notch range by setting $\omega_{\min}=5,000$ and $\omega_{\max}=8,000$ Hz. Signal-to-noise ratio range is again set to $8$-$32$ dB. 

This augmentation scheme modifies the input signal via two convolutional operators (steps $3$ and $4$). These are linear operators that can be realized by multiplying the input signal with a circulant matrix. In particular, there exist circulant matrices $C_0$ and $C_{\omega}$ that correspond to notch filters $h_0$ and $h_{\omega}$ (respectively) such that given an input signal $x$, the corresponding transformed signal is $z = C_{\omega}C_0x$. In terms of Eq.~\ceq{eq:gauss-mixture}, this means that each of the frequency dips $\cbrackets{\omega_k}_{k=1}^p$ defines a mixture component such that $\mu_{ik}\brackets{x_i} \coloneqq \mu_{\omega_k}\brackets{x_i}=C_{\omega_k}C_0x_i$ with $1\leq i \leq n$ and $1\leq k \leq p$, and where $x_i$ denotes a training sample. The variance term in this augmentation scheme is isotropic and it accounts for the signal-to-noise ratio.

\subsection{Wide Band-pass Filtered Signals}
\label{subsec:bandpass}

Spurious correlations due to microphone effects are challenging to address when learning directly in the waveform domain. The main goal in this section is to devise an augmentation scheme that could be effective in dealing with adverse conditions related to mismatch between training and test  microphones. Such effects can be characterized by suppression of  content in high and low frequency regions of the spectrum. Hence, we would like to pass the input signal through a wide band-pass filter that removes information at low and high frequencies, and replace that information by white noise, which is independent of labels, and thus biases the learning process away from over-fitting the microphone effects.

\begin{algorithm}[t]
	\algsetup{linenosize=\tiny}
	\caption{\textsc{noisy widepass}} 
	\label{alg:noisy-widepass}
	\begin{algorithmic}[1]
		{
			\fontsize{9}{11}\selectfont
			\REQUIRE 
			audio signal $x\brackets{t}$, sampling rate $f$, bandpass frequency range given by $\omega_{\min}$ and $\omega_{\max}$, filterbank size $p$, filter support size $T$, \textsc{snr} range given by $\gamma_{\min}$ and $\gamma_{\max}$\vspace*{1ex}  
			\STATE $\{\omega_i\}_{i=1}^p \leftarrow \textsc{evenly\_spaced\_modes} \brackets{\omega_{\min}, \omega_{\max}, f, p}$\vspace*{0.5ex}
			\STATE $\{\xi_i\}_{i=1}^p \leftarrow \textsc{wide\_bandwidths} \brackets{\{\omega_i \}_{i=1}^p, \omega_{\min}, \omega_{\max}, f}$\vspace*{0.5ex}
			\STATE $(\omega, \xi) \sim \mathcal{U}_{\cbrackets{(\omega_1, \xi_1), \dots, (\omega_p, \xi_p)}}$\vspace*{0.5ex}
			\STATE $h\brackets{t} \leftarrow \textsc{parzen\_filter}\brackets{\omega, \xi, T}$\vspace*{0.5ex}
			\STATE $z\brackets{t} \leftarrow \brackets{x * h}\brackets{t}$ and $\epsilon(t) \sim \mathcal{N}(0, 1)$\vspace*{0.5ex}
			\STATE $\epsilon\brackets{t} \leftarrow \textsc{snr\_scale} \brackets{\epsilon\brackets{t}, \gamma}$  with $\gamma \sim \mathcal{U}\brackets{\gamma_{\min},\gamma_{\max}}$\vspace*{0.5ex}
			\STATE \textbf{return} $z\brackets{t}+\epsilon\brackets{t}$
		}
	\end{algorithmic}
\end{algorithm}

Algorithm~\ref{alg:noisy-widepass} provides a pseudo-code description of this signal transformation. In the first step, we create a set of $p$ evenly spaced modulation frequencies covering the frequency range provided as input. Then, the algorithm creates a set of $p$ bandwidths (step $2$) such that the resulting filters are with wide band-pass properties (see step $4$). Following this, we select one of these filters uniformly at random (steps $3$ and $4$) and convolve it with the input signal (step $5$), thus retaining only the information covered by the filter support. In the final step, a white noise vector is scaled such that the additive corruption of band-passed signal has the appropriate signal-to-nose ratio, sampled uniformly from the range given as input to the algorithm (step $6$).
We simulate this algorithm with $p=8$ bandpass Parzen filters (see Section~\ref{sec:parznets}) with the support over frequency range $50$-$7950$ Hz. The bandwidths are selected to follow the Mel-scale, that is, the bandwidth $\xi_i$ of the filter with centre frequency $\omega_i$ is set to be approximately equal to the width of the band at that same frequency on the Mel-scale. In the case of relatively small number of filters, that ensures that filters have wider bandwidths. For the signal-to-noise ratio we use the range of $8$-$32$ dB.

This augmentation scheme alters the input signal via convolution with a band-pass filter. The operation can be realized by multiplication with a circulant matrix, i.e., for a band-pass filter $h$ there exists a circulant matrix $C_h$ such that $z=C_{h}x$. In terms of Eq.~\ceq{eq:gauss-mixture}, each of the band-pass filters $ \cbrackets{(\omega_k, \xi_k)}_{k=1}^p$ from steps $1$-$4$ in Algorithm~\ref{alg:noisy-widepass} defines a mixture component such that $\mu_{ik}\brackets{x_i} \coloneqq \mu_{\omega_k, \xi_k}\brackets{x_i}=C_{\omega_k, \xi_k}x_i $ with $1\leq i \leq n$ and $1\leq k \leq p$, and where $x_i$ denotes a training sample. As in the previous scheme, the variance term is isotropic and accounts for the signal-to-noise ratio.

\subsection{Reverberation Effects}
\label{subsec:rir}

This section covers a data augmentation scheme based on room impulse response modeling. Reverberations introduce spurious correlations in speech signals as the result of a large distance between speakers and a microphone~\cite{Yoshioka12}. In particular, speech signal reaches  a microphone via a direct line-of-sight path, followed by first, second and higher order reflections  off the  walls and other objects. These reflections, referred to as reverberation,  can be represented as linear convolutions of the speech signal with the room impulse response. There are two stages characteristic to this process, early reflections that typically occur within $50$ ms, followed by a dense series of higher-order reflections called late reverberations~\cite{Yoshioka12,reverbaugment}. The challenging aspect of this process is the fact that late reverberation is non-stationary and is, thus, not addressable by standard noise compensation techniques such as vector Taylor series~\cite{Yoshioka12}. The aim of the augmentation scheme introduced in this section is to scramble spurious correlations that can occur as a result or distant-talking and reverberations. \emph{Thus, we are not interested in designing noise compensation mechanisms but are willing to delegate this task to the network itself with inductive bias guided by examples of linear convolutions of input signals with different room impulse responses.}

\begin{algorithm}[t]
	\algsetup{linenosize=\tiny}
	\caption{\textsc{noisy rir}} 
	\label{alg:noisy-rir}
	\begin{algorithmic}[1]
		{
			\fontsize{9}{11}\selectfont
			\REQUIRE 
			audio signal $x\brackets{t}$, sampling rate $f$, set of \textsc{3d} room configurations $\mathcal{C}$, set of wall materials $\mathcal{M}$, source to microphone distances given by $d_{\min}$ and $d_{\max}$, \textsc{snr} range given by $\gamma_{\min}$ and $\gamma_{\max}$\vspace*{1ex}  
			\STATE $c \sim \mathcal{U}_{\cbrackets{c_1, \dots, c_{\absolute{\mathcal{C}}}}}$ with $ \mathcal{C}=\{c_i\}_{i=1}^{\absolute{\mathcal{C}}}$\vspace*{0.5ex}
			\STATE $m \sim \mathcal{U}_{\cbrackets{m_1, \dots, m_{\absolute{\mathcal{M}}}}}$ with $\mathcal{M}=\{m_i\}_{i=1}^{\absolute{\mathcal{M}}}$\vspace*{0.5ex}
			\STATE $\mathrm{mic} \leftarrow \mathrm{vec}\brackets{u_1,u_2,u_3}$ with $u_i \sim \mathcal{U}(0, c\sbrackets{i})$\vspace*{0.5ex}
			\STATE $h \leftarrow \textsc{rir} \brackets{c, m, \mathrm{mic}, d}$ with $d \sim \mathcal{U}\brackets{d_{\min}, d_{\max}}$\vspace*{0.5ex}
			\STATE $z\brackets{t} \leftarrow \brackets{x * h}\brackets{t}$ and $\epsilon(t) \sim \mathcal{N}(0, 1)$\vspace*{0.5ex}
			\STATE $\epsilon\brackets{t} \leftarrow \textsc{snr\_scale} \brackets{\epsilon\brackets{t}, \gamma}$  with $\gamma \sim \mathcal{U}\brackets{\gamma_{\min},\gamma_{\max}}$\vspace*{0.5ex}
			\STATE \textbf{return} $z\brackets{t}+\epsilon\brackets{t}$
		}
	\end{algorithmic}
\end{algorithm}

Algorithm~\ref{alg:noisy-rir} provides a pseudo-code description of this augmentation scheme. First, we select (uniformly at random) one of the possible room configurations (step $1$) and a wall material setting (step $2$). Following this, the algorithm samples a location of the microphone based on the selected room dimensions (step $3$) along with the distance to an acoustic source (step $4$). The selected parameters are then passed to the \emph{pyroomacoustics} library~\cite{pyroomacoustics}, which generates a room impulse response. The algorithm convolves the input speech signal with the selected room impulse response and further corrupts the resulting signal with an additive white noise, appropriately scaled to respect the selected signal-to-noise ratio (steps $5$-$6$). The reason for adding white noise to the reverberated signal is to avoid having repeated noise types in the training samples as deep learning models may learn to remove them, defeating the purpose of scrambling spurious correlations introduced by distant-talking. We note here that white noise has been previously combined with reverberated signals in~\cite{noisy-rir}.

In our implementation of this data augmentation scheme, we use the following three room configurations (in meters)
\begin{align*}
\mathcal{C}=\cbrackets{[4,4,2.5], [10,10,3.5], [2.5, 1.5, 1.5]}
\end{align*}
along with five types of wall materials $\mathcal{W}=\{$\emph{hard surface, marble floor, wooden door, glass window, hairy carpet}$\}$. For each of the material types, we select one of the four possible scattering modes~\cite{pyroomacoustics}: \emph{none, rpg-skyline, classroom tables, and rectangular prism boxes}. The distance to the microphone is selected uniformly at random in the range between $3$ cm and $3$ m. As in previous cases, white noise is added with the resulting signal-to-noise ratio in the range $8$-$32$ dB.

Similar to the two previously covered augmentation schemes, the input signal is altered using the convolution operator defined by a \textsc{rir} filter. In the context of Eq.~\ceq{eq:gauss-mixture} and our theoretical analysis this means that each \textsc{rir} filter corresponds to a mixture component. More formally, for a \textsc{rir} filter $h$ there exists a circulant matrix $C_{h}$ such that $z=C_{h}x$. Thus, mixture component means (see Eq.~\ref{eq:gauss-mixture}) that arise as a result of using this augmentation scheme are given by $\mu_{ik}\brackets{x_i} \coloneqq \mu_{h_k}\brackets{x_i}=C_{h_{k}}x_i$ with $1 \leq i \leq n$ and a set of \textsc{rir} filters $\cbrackets{h_k}_k$.

\section{Related Work}
\label{sec:rw} 

Data augmentation has been widely used in the past to improve performance of acoustic models on different tasks. The process of augmenting the training sample is typically done in an improvised manner using well established heuristics (e.g., masking blocks of frequency channels, speed and tempo perturbations, additive noise, etc.) and little understanding of the underlying machine learning principles. The latter aspect is a contribution  of our work compared to previously devised augmentation schemes for improving robustness to distribution shifts and out-of-distribution generalization performance (e.g., see Section~\ref{sec:theory}). Previous data augmentation approaches can be divided into three classes: \emph{i}) acoustic data perturbations, \emph{ii}) text-to-speech augmentation, and \emph{iii}) augmentation with unsupervised data (semi-supervised and self-supervised learning).

The first class of approaches aims at reducing the divergence between training and test conditions/distributions by transformations of the inputs that tackle common sources of mismatch such as additive and channel noise, reverberations, microphone types, speaker variations, etc. The perturbations are typically label-preserving and do not require additional alignments (e.g., when learning with hybrid \textsc{hmm-dnn} models). This process of adding label-preserving noisy data with different types and levels of noise to the original training samples is known as \emph{multi-condition/style training}~\cite[][]{multistyle} and it is used frequently for learning of robust acoustic models~\cite[e.g., see][]{yu2013feature,Cui15,aurora4}. When the noisy data is generated using a database with a limited number of naturally occurring environmental noise signals the approach may fail to generalize beyond the training distribution. More specifically, it can happen that a neural network \emph{memorizes} all the noise signals and learns to \emph{subtract} them from the augmented data~\cite[e.g., see][]{deepspeech}. Indeed, the noise in multi-condition training is typically constant --- the utterances are corrupted prior to training and frames are repeatedly (in each epoch) presented to a learning algorithm with identical corruptions. Hence, it can be challenging for the network to differentiate between signal and spurious correlation because they might be coupled throughout the training process. It is for this reason that our augmentation schemes (presented in Section~\ref{sec:data-augmentation}) rely on additive Gaussian noise with various signal-to-noise ratios to generate diverse label-preserving input perturbations. Moreover, our augmentation approaches are amenable to \emph{online} data perturbations where at the onset of an epoch one can randomly select an additive corruption or proceed with the original utterance (see Sections~\ref{subsec:fbank} and~\ref{subsec:ami-sdm}). 
\textsc{SpecAugment}~\cite{specaugment} is a recently proposed highly effective data augmentation scheme from this class. It operates on the log-mel-spectrogram of the input audio and consists of steps such as feature warping, masking blocks of frequency channels, and masking blocks of time steps. A shortcoming is the restriction to filterbank features, which precludes its use in other types of acoustic models (e.g., truly end-to-end speech recognition systems). In contrast to this, the augmentation schemes proposed in Section~\ref{sec:data-augmentation} apply to all types of acoustic models and to some extent model the policies characteristic to~\cite{specaugment}. In particular, Algorithm~\ref{alg:noisy-widepass} transforms the input signal and replaces blocks of frequency channels with white noise which is independent of labels and can, thus, scramble spurious correlations in that part of the spectrum. Related to our augmentation schemes (i.e., Algorithm~\ref{alg:bandlimited}) is the bandpass noise generation approach proposed in~\cite{banpassaug}. More specifically, that approach takes a database of naturally occurring sounds along with a bandpass filterbank to generate a set of bandlimited noise signals.  
The empirical results in~\cite{banpassaug} report a $7\%$ relative improvement in generalization to unseen noise conditions (far-field device and car noise). In contrast to this approach, Algorithm~\ref{alg:bandlimited} relies exclusively on independent Gaussian samples that scramble spurious correlations in the low-frequency region. Another set of transformations directly relevant for our work are room impulse response (\textsc{rir}) perturbations. Previous work reports improvement in distant and noisy speech as a result of this augmentation scheme~\cite[e.g., see][]{Yoshioka12,reverbaugment}. We extend this line of work by combining additive Gaussian noise with \textsc{rir} perturbation, thus tackling the potential memorization effects when learning with an over-parameterized neural network (see Algorithm~\ref{alg:noisy-rir}). Speed and tempo perturbations are also frequently used transformations of the input signal that can improve generalization in conversational speech~\cite[e.g., see][]{speedtempo}. A potential issue with this approach is the change in the duration of an utterance, requiring re-alignment of target labels (e.g., senones in hybrid \textsc{hmm-dnn} models) for a transformed audio that might not be supported by the training distribution (thus resulting in poor/noisy training labels). Here it is important to note that speed perturbations were designed to emulate the vocal tract length perturbations (\textsc{vtlp}) that aim at improving robustness by adding synthetic speaker variations~\cite[e.g., see][]{speedtempo,vocaltract}. Interestingly, it has been observed in~\cite{vocaltract} that the most effective perturbations are those applied to linear spectrograms, which is in line with the theoretical principles outlined in~\cite[Section II.C,][]{varparznets} and realized by the augmentation schemes proposed in Section~\ref{sec:data-augmentation}. Namely, in all of the cases the transformations of the inputs aim at removing spurious correlations present in linear spectrograms.

The second class of approaches aims at improving robustness of acoustic models by synthetically adding data using speech synthesis. This allows for training of models with utterances that were not part of the original training sample. Thus, the ultimate goal of these approaches is not to achieve robustness to novel acoustic conditions but to provide further flexibility when it comes to language constructions. Prominent augmentation schemes from this class are the approaches proposed by~\cite{Laptev20} and~\cite{Renduchintala18}. A common issue with this class of approaches is the lack of high quality training data (i.e., audio recordings), which manifests itself especially in low-resource languages. Another issue is that for many low-resource languages it may be difficult to find textual data in the amount typically used for building language models (e.g., for English or Mandarin). 

The third class of approaches aims at improving robustness by leveraging large amounts of unlabeled data using semi-supervised and self-supervised learning. For example, in~\cite{Ragni2014DataAF} an approach for acoustic models operating with low-resource languages has been devised that combines semi-supervised learning and vocal tract length perturbations. Perhaps the most prominent approach from this class of learning algorithms is the group of neural architectures known as \textsc{wav2vec}~\cite{schneider2019wav2vec,baevski2020wav2vec}, which rely on self-supervised learning and large amounts of unlabeled data to extract robust representations of sub-phonetic units. Typically, the learned representation are adapted to the task at hand with additional training using a small amount of labeled data. The main advantage of this class of approaches is in the fact that unlabelled data is typically easy to obtain, e.g., news broadcasts covering various different speaker and noise conditions~\cite{Ragni2014DataAF}. A potential shortcoming of \emph{unsupervised} approaches can be the lack of high quality transcriptions, which are important for transferring the learned representations to the task at hand. In~\cite{Ragni2014DataAF} it has been observed that the lack of transcriptions can limit gains from some semi- and unsupervised approaches such as those based on discriminative training~\cite{galeswoodland} and speaker adaptations based on discriminative criteria~\cite{woodlanddiscrim}. Recent empirical results, however, show that self-supervised learning can be effective in low-resource \textsc{asr}~\cite{schneider2019wav2vec,baevski2020wav2vec}.

\section{Experiments}
\label{sec:exps}

The ultimate goal of our empirical evaluation is to assess the effectiveness of the proposed augmentation techniques (Section~\ref{sec:data-augmentation}) and underlying theoretical principles (Section~\ref{sec:theory}) in a controlled setting with a significant mismatch between training and test conditions. To this end, we perform a series of experiments on \textsc{aurora4}~\cite{aurora4} and \textsc{ami}~\cite{renals2007recognition} datasets with the help of the corresponding Kaldi recipes~\cite{kaldi}. 

\textsc{aurora4} is a well known benchmark dataset that has been assembled for the purpose of evaluating the effectiveness of acoustic models in out-of-distribution generalization. In particular, the benchmark includes a Kaldi recipe/setting known as \emph{clean-condition training} with a set of clean speech training utterances recorded using a high-quality close-talking \emph{Sennheiser} microphone and a number of test utterances recorded using one of $18$ different microphones. While some of the test utterances are recorded in clean conditions, the majority of them are noisy and cover various different noise conditions such as street traffic, train station, car, babble, restaurant, and airport. The signal-to-noise ratio (\textsc{snr}) in these test recordings ranges from $5$ to $15$ dB. Hence, the clean-condition Kaldi recipe for \textsc{aurora4} allows for assessing two modes of out-of-distribution generalization: \emph{i}) robustness to different types and levels of noise unseen during training, and \emph{ii}) robustness relative to bias introduced by microphones and recording devices. In addition to this, the benchmark also includes a Kaldi recipe/setting known as \emph{multi-condition training} that has been designed such that the conditions (i.e., noise types and microphones) present in training match those for the test utterances. More specifically, the training sample in this setting includes recordings from Sennheiser and other microphones, and it covers various different noise types~\cite{aurora4}. An important difference between acoustic conditions present in the two samples for this setting is in the range of signal-to-noise ratios ($10$-$20$ dB in training vs $5$-$15$ dB in test utterances). Here, it is important to note that throughout our empirical evaluation we maintain the mismatch in the signal-to-noise ratios between training data (detailed below) and the test conditions of \textsc{aurora4}. 

In our augmentation schemes, we are not relying on any of the naturally occurring noise types such as the ones present in test samples of \textsc{aurora4}. Thus, a statistically significant improvement over clean-condition training, measured via the word error rate (\textsc{wer}) on unseen noisy tests samples, would imply that the proposed approach generalizes to unseen acoustic conditions. The alternative multi-condition training recipe allows us to learn a competitive baseline model with an identical architecture and can, thus, be used to assess the potential of the proposed augmentation schemes and underlying principles for learning noise-robust acoustic models.

The second benchmark dataset \textsc{ami} comes with different Kaldi recipes for data recorded using headset and distant microphones. Again, we illustrate the utility of the proposed algorithms in out-of-distribution generalization, this time generalizing from headset recorded to distant-talking speech.

In all of our experiments, we train a context dependent model based on frame labels (i.e., \textsc{hmm} state ids) generated using a triphone model from Kaldi~\cite{kaldi} with $25$ ms frames and $10$ ms stride. The data splits (training/development/evaluation) are identical to the ones from the corresponding Kaldi recipes. In the preprocessing step, we assign the Kaldi frame label to a $200$ ms long segment of raw speech centered at the original Kaldi frame. The Parzen convolution block is initialized by taking the modulation frequencies to be equidistant in mel-scale. The bands of filters are initialized as in \textsc{fbank} features. For convolutional and dense blocks in our network, we employ the Xavier initialization scheme~\cite{xavier} with magnitude $0.005$. While the convolutional blocks are initialized with the factor type \emph{in}, the dense blocks use the \emph{avg} type. The feature extraction layers (i.e., Parzen and convolutional parameters) are updated using the \textsc{rmsprop} optimizer with the initial learning rate set to $0.0008$. The multi-layer perceptron blocks are updated using stochastic gradient descent with the initial learning rate set to $0.08$. A similar combination of optimizers (all network parameters are optimized jointly) was used in~\cite{sincnet,parznets20}. When training using data augmentation, we decrease the learning rates by a factor of $2$ at the end of an epoch, apart from the two initial epochs, and repeat this until the completion of training. We use minibatches of $512$ samples and terminate the training process after $8$ epochs. When training in the original clean- and multi-condition settings, we terminate the training process after $25$ epochs and decrease the learning rates by a factor of $2$ at the end of an epoch if the relative improvement in the classification accuracy on the validation fold is below $0.1\%$.

\begin{table*}[t]
	\centering\fontsize{7}{9}\selectfont  
	\caption{The table reports the word error rates obtained on \textsc{aurora4} using the Kaldi clean-condition recipe. To assess the effectiveness of the proposed augmentation schemes, the training fold has been replicated by Algorithms~\ref{alg:bandlimited}-\ref{alg:noisy-rir}. 
	The relative performance degradation/improvement is computed with respect to the second column (colored in orange), which includes all replicas of the training fold.
    The symbol $-$ implies that the corresponding replica of the clean-condition training fold has been excluded from the training set in that simulation.  
	In the rightmost column we report the results using the training set from the Kaldi clean-condition recipe alone. The column labelled \textsc{detr.} reports the relative deterioration in the word error rate as a result of removing the contribution of an augmentation scheme from the training recipe --- the entries colored in red reflect the negative effects on the performance as a result of training without that particular vicinal density characterization.}
	\begin{tabular}[t]{l|c|r|r|r|r|r|r|r|r|r|r|r|r}
		\multirow{2}{*}{\textsc{test sample}}  
		&  \cellcolor{orange!10} \textsc{\thead[l]{\fontsize{6}{8}\selectfont clean \\ \fontsize{6}{8}\selectfont bandlim. \\ \fontsize{6}{8}\selectfont notch \\ \fontsize{6}{8}\selectfont widepass \\ \fontsize{6}{8}\selectfont rir \\ \fontsize{6}{8}\selectfont gauss}}
		& \multicolumn{2}{c|}{\textsc{\thead[l]{\fontsize{6}{8}\selectfont clean \\ \fontsize{6}{8}\selectfont - \\ \fontsize{6}{8}\selectfont notch \\ \fontsize{6}{8}\selectfont widepass \\ \fontsize{6}{8}\selectfont rir \\ \fontsize{6}{8}\selectfont gauss}}}
		& \multicolumn{2}{c|}{\textsc{\thead[l]{\fontsize{6}{8}\selectfont clean \\ \fontsize{6}{8}\selectfont bandlim. \\ \fontsize{6}{8}\selectfont - \\ \fontsize{6}{8}\selectfont widepass \\ \fontsize{6}{8}\selectfont rir \\ \fontsize{6}{8}\selectfont gauss}}}  
		& \multicolumn{2}{c|}{\textsc{\thead[l]{\fontsize{6}{8}\selectfont clean \\ \fontsize{6}{8}\selectfont bandlim. \\ \fontsize{6}{8}\selectfont notch \\ \fontsize{6}{8}\selectfont - \\ \fontsize{6}{8}\selectfont rir \\ \fontsize{6}{8}\selectfont gauss}}} 
		& \multicolumn{2}{c|}{\textsc{\thead[l]{\fontsize{6}{8}\selectfont clean \\ \fontsize{6}{8}\selectfont bandlim. \\ \fontsize{6}{8}\selectfont notch \\ \fontsize{6}{8}\selectfont widepass \\ \fontsize{6}{8}\selectfont - \\ \fontsize{6}{8}\selectfont gauss}}}
		& \multicolumn{2}{c|}{\textsc{\thead[l]{\fontsize{6}{8}\selectfont clean \\ \fontsize{6}{8}\selectfont bandlim. \\ \fontsize{6}{8}\selectfont notch \\ \fontsize{6}{8}\selectfont widepass \\ \fontsize{6}{8}\selectfont rir \\ \fontsize{6}{8}\selectfont -}}} 
		&  \multicolumn{2}{c|}{\textsc{\thead{\fontsize{6}{8}\selectfont clean \\ \fontsize{6}{8}\selectfont  - \\\fontsize{6}{8}\selectfont   -\\\fontsize{6}{8}\selectfont   -\\\fontsize{6}{8}\selectfont  - \\\fontsize{6}{8}\selectfont   -}}}  \\\cline{2-14}
		& \textsc{\fontsize{6}{8}\selectfont err.} & \textsc{\fontsize{6}{8}\selectfont err.~} & \textsc{\fontsize{6}{8}\selectfont detr.} & \textsc{\fontsize{6}{8}\selectfont err.~} & \textsc{\fontsize{6}{8}\selectfont detr.} & \textsc{\fontsize{6}{8}\selectfont err.~} & \textsc{\fontsize{6}{8}\selectfont detr.} & \textsc{\fontsize{6}{8}\selectfont err.~} & \textsc{\fontsize{6}{8}\selectfont detr.} & \textsc{\fontsize{6}{8}\selectfont err.~} & \textsc{\fontsize{6}{8}\selectfont detr.} &
		\textsc{\fontsize{6}{8}\selectfont err.~} &
		\textsc{\fontsize{6}{8}\selectfont detr.}\\
		\hline	
		\multicolumn{14}{l}{\cellcolor{blue!5} \textsc{a. summary over clean speech with training microphones}}\\
		\hline
		\textsc{clean} & \cellcolor{orange!10} $2.58$ & $2.58$ & $-$ & $2.56$ & \cellcolor{green!25} $1\%$ & $2.58$ & $-$ & $2.47$ & \cellcolor{green!25} $4\%$ & $2.54$ & \cellcolor{green!25} $2\%$  & $\mathbf{1.96}$ & \cellcolor{green!25} $24\%$ \\
		
		\hline
		\multicolumn{14}{l}{\cellcolor{blue!5} \textsc{b. summary over noisy speech with training microphones}}  \\
		\hline
		\textsc{car} &  \cellcolor{orange!10} $3.53$ & $3.61$ & \cellcolor{red!25} $2\%$ & $4.04$ & \cellcolor{red!25} $14\%$ & $3.92$ & \cellcolor{red!25} $11\%$ & $3.89$ & \cellcolor{red!25} $10\%$ & $\mathbf{3.36}$ & \cellcolor{green!25} $5\%$ & $15.92$ & \cellcolor{red!25} $351\%$ \\
		
		\textsc{babble} &  \cellcolor{orange!10} $6.58$ & $9.86$ & \cellcolor{red!25} $50\%$ & $5.98$ & \cellcolor{green!25} $9\%$ & $\mathbf{5.66}$ & \cellcolor{green!25} $14\%$ & $6.58$ & $-$ & $5.74$ & \cellcolor{green!25} $13\%$ & $16.66$ & \cellcolor{red!25} $153\%$\\
		
		\textsc{restaurant} &  \cellcolor{orange!10} $7.64$ & $8.16$ & \cellcolor{red!25} $7\%$ & $7.64$ & $-$ & $\mathbf{7.38}$ & \cellcolor{green!25} $3\%$ & $8.41$ & \cellcolor{red!25} $10\%$ & $7.45$ & \cellcolor{green!25} $2\%$ & $15.99$ & \cellcolor{red!25} $109\%$\\
		
		\textsc{street} &  \cellcolor{orange!10} $\mathbf{7.04}$ & $7.23$ & \cellcolor{red!25} $3\%$ & $8.43$ & \cellcolor{red!25} $20\%$ & $7.83$ & \cellcolor{red!25} $11\%$ & $8.14$ & \cellcolor{red!25} $16\%$ & $7.32$ & \cellcolor{red!25} $4\%$ & $25.67$ & \cellcolor{red!25} $265\%$\\
		
		\textsc{airport} &  \cellcolor{orange!10} $\mathbf{6.26}$ & $9.21$ & \cellcolor{red!25} $47\%$ & $6.41$ & \cellcolor{red!25} $2\%$ & $6.44$ & \cellcolor{red!25} $3\%$ & $6.46$ & \cellcolor{red!25} $3\%$ & $6.63$ & \cellcolor{red!25} $6\%$ & $12.67$ & \cellcolor{red!25} $102\%$\\
		
		\textsc{train}  &  \cellcolor{orange!10} $7.06$ & $\mathbf{6.65}$ & \cellcolor{green!25} $6\%$ & $8.16$ & \cellcolor{red!25} $16\%$ & $7.73$ & \cellcolor{red!25} $9\%$ & $8.89$ & \cellcolor{red!25} $26\%$ & $7.29$ & \cellcolor{red!25} $3\%$ & $26.56$ & \cellcolor{red!25} $276\%$\\
		
		\hline
		\textsc{b. average}  &  \cellcolor{orange!10} $6.35$ & $7.45$ & \cellcolor{red!25} $17\%$ & $6.78$ & \cellcolor{red!25} $7\%$ & $6.49$ & \cellcolor{red!25} $2\%$ & $7.06$ & \cellcolor{red!25} $11\%$ & $\mathbf{6.30}$ & \cellcolor{green!25} $1\%$ & $18.91$ & \cellcolor{red!25} $198\%$\\
		
		\hline
		\multicolumn{14}{l}{\cellcolor{blue!5} \textsc{c. summary over clean speech with different microphones (unseen during training)}}\\
		\hline
		\textsc{clean} &  \cellcolor{orange!10} $7.79$ & $7.58$ & \cellcolor{green!25} $3\%$ & $8.44$ & \cellcolor{red!25} $8\%$ & $12.29$ & \cellcolor{red!25} $58\%$ & $\mathbf{7.49}$ & \cellcolor{green!25} $4\%$ & $8.03$ & \cellcolor{red!25} $3\%$  & $19.47$ & \cellcolor{red!25} $150\%$\\
		
		\hline
		\multicolumn{14}{l}{\cellcolor{blue!5} \textsc{d. summary over noisy speech with different microphones (unseen during training)}}\\
		\hline
		\textsc{car} &  \cellcolor{orange!10} $10.46$ & $\mathbf{7.85}$ & \cellcolor{green!25} $25\%$ & $12.46$ & \cellcolor{red!25} $19\%$ & $15.32$ & \cellcolor{red!25} $46\%$ & $9.98$ & \cellcolor{green!25} $5\%$ & $8.78$ & \cellcolor{green!25} $16\%$ & $34.32$ & \cellcolor{red!25} $228\%$\\
		
		\textsc{babble} &  \cellcolor{orange!10} $16.22$ & $17.60$ & \cellcolor{red!25} $9\%$ & $16.59$ & \cellcolor{red!25} $2\%$ & $19.04$ & \cellcolor{red!25} $17\%$ & $18.70$ & \cellcolor{red!25} $15\%$ & $\mathbf{15.37}$ & \cellcolor{green!25} $5\%$ & $38.93$ & \cellcolor{red!25} $140\%$\\
		
		\textsc{restaurant} &  \cellcolor{orange!10} $18.16$ & $\mathbf{18.08}$ & $-$ & $19.63$ & \cellcolor{red!25} $8\%$ & $21.65$ & \cellcolor{red!25} $19\%$ & $20.79$ & \cellcolor{red!25} $14\%$ & $18.79$ & \cellcolor{red!25} $3\%$  & $36.52$ & \cellcolor{red!25} $101\%$\\
		
		\textsc{street} &  \cellcolor{orange!10} $18.74$ & $\mathbf{16.63}$ & \cellcolor{green!25} $ 11\%$ & $20.74$ & \cellcolor{red!25} $11\%$ & $22.73$ & \cellcolor{red!25} $21\%$ & $20.51$ & \cellcolor{red!25} $9\%$ & $17.21$ & \cellcolor{green!25} $8\%$ & $49.32$ & \cellcolor{red!25} $163\%$\\
		
		\textsc{airport} &  \cellcolor{orange!10} $\mathbf{16.50}$ & $17.37$ & \cellcolor{red!25} $5\%$ & $16.98$ & \cellcolor{red!25} $3\%$ & $20.31$ & \cellcolor{red!25} $23\%$ & $18.57$ & \cellcolor{red!25} $13\%$ & $16.66$ & \cellcolor{red!25} $1\%$ & $35.59$ & \cellcolor{red!25} $116\%$\\
		
		\textsc{train} &  \cellcolor{orange!10} $19.15$ & $\mathbf{16.44}$ & \cellcolor{green!25} $14\%$ & $20.94$ & \cellcolor{red!25} $9\%$ & $22.92$ & \cellcolor{red!25} $20\%$ & $21.32$ & \cellcolor{red!25} $11\%$ & $18.48$ & \cellcolor{green!25} $3\%$ & $48.07$ & \cellcolor{red!25} $151\%$\\
		
		\hline
		\textsc{d. average}  &  \cellcolor{orange!10} $16.54$ & $\mathbf{15.66}$ & \cellcolor{green!25} $5\%$ & $17.89$ & \cellcolor{red!25} $8\%$ & $20.33$ & \cellcolor{red!25} $23\%$ & $18.31$ & \cellcolor{red!25} $11\%$ & $15.88$ & \cellcolor{green!25} $4\%$ & $40.46$ & \cellcolor{red!25} $145\%$\\
		
		\hline
		\multicolumn{14}{l}{\cellcolor{blue!5} \textsc{summary over all $14$ test samples}}\\
		\hline
		\textsc{average} &  \cellcolor{orange!10} $10.55$ & $10.63$ & \cellcolor{red!25} $1\%$ & $11.36$ & \cellcolor{red!25} $8\%$ & $12.56$ & \cellcolor{red!25} $19\%$ & $11.59$ & \cellcolor{red!25} $10\%$ & $\mathbf{10.26}$ & \cellcolor{green!25} $3\%$  & $26.98$ & \cellcolor{red!25} $156\%$\\
		\hline
		
		\hline
	\end{tabular}
	\label{tbl:aurora4}\vspace{-1ex}
\end{table*}

\subsection{Impact of the Augmentation Schemes on Robustness}
\label{subsec:impact}

In the first set of experiments, the goal is to estimate the impact of individual augmentation schemes (Algorithms~\ref{alg:bandlimited}-~\ref{alg:noisy-rir}) on the robustness to unseen noise conditions. To this end, we generate a transformed set of training utterances from the Kaldi clean-condition recipe using each of the augmentations schemes, along with a set corrupted using the Gaussian noise alone. The reason for including the latter corruption is to establish whether there are spurious correlations that are complementary to the proposed augmentation schemes and addressable via plain additive Gaussian noise. We refer to the utterances produced by Algorithms~\ref{alg:bandlimited}-~\ref{alg:noisy-rir} (respectively) as \textsc{bandlimited}, \textsc{notch}, \textsc{widepass}, and \textsc{rir}. In the first experiment, we train the \textsc{parznets 2d} model~\cite{parznets20} using all the available data (\textsc{clean}, Algorithms~\ref{alg:bandlimited}-~\ref{alg:noisy-rir}, and additive \textsc{gauss} perturbations). The second column in Table~\ref{tbl:aurora4} summarizes the performance of the model across different test sets (\textsc{a}: clean speech with training microphones, \textsc{b}: noisy speech with training microphones, \textsc{c}: clean speech with different microphones, and \textsc{d}: noisy speech with different microphones). In the remainder of this subsection, we will use the word error rate obtained in this way (i.e., \textsc{wer} $10.55\%$) as a reference and quantify the influence of an augmentation scheme relative to it by training the identical model without the corresponding training sample contribution.

\subsubsection*{Algorithm~\ref{alg:bandlimited} (\textsc{bandlim.})} The third column in Table~\ref{tbl:aurora4} summarizes the results obtained by removing the contribution of this augmentation scheme from the training sample. The sub-column labeled as \textsc{err.} reports the word error rate that has slightly increased ($1\%$ relative) as a result of this intervention. The augmentation scheme contributes to significant improvement (colored in red) in word error rate on noisy samples recorded using the training microphones (babble and airport noise, $50\%$ and $47\%$ relative). This comes at the expense of picking up spurious correlations that can be associated with microphones, resulting in performance degradation on sample \textsc{D} (car, train, and street noise are affected the most). Overall, the augmentation scheme can help with spurious correlations specific to some types of additive noise.

\subsubsection*{Algorithm~\ref{alg:double-notch} (\textsc{notch})} The fourth column in Table~\ref{tbl:aurora4} quantifies the performance for this setting, with the notch contribution removed from the training sample. Overall, there is an increase in the word error rate which signifies the importance of this augmentation scheme ($8\%$ relative over the whole test fold). The augmentation scheme can help with spurious correlations introduced by all the noise types considered, except for babble noise. Moreover, the improvement in robustness is consistent and independent of the microphones used.

\subsubsection*{Algorithm~\ref{alg:noisy-widepass} (\textsc{widepass})} The fifth column in Table~\ref{tbl:aurora4} summarizes the results 
with the \textsc{widepass} contribution removed from the training sample. The empirical results indicate that this is the most effective augmentation scheme for dealing with spurious correlations (removal results in increased word error rate, $19\%$ relative). When it comes to dealing with microphone effects, this is the most effective strategy among the ones considered, with $58\%$ and $23\%$ relative improvement on clean and noisy speech, respectively. In comparison to other sampling schemes, \textsc{widepass} augmentation is extremely effective in dealing with spurious correlations attributed to noisy speech with different microphones (see the results for sample \textsc{d}).

\begin{table*}[t]
	\centering\fontsize{7}{9}\selectfont  
	\caption{The word error rates (\%) obtained on different test sets of \textsc{aurora4} with various training settings.}
	\begin{tabular}[t]{l|c|c|c|c|c|c|c|c|c|c}
		\textsc{training setting~\textbackslash~input type}&\multicolumn{6}{c|}{\textsc{\fontsize{6}{7}\selectfont raw speech}} & \multicolumn{4}{c}{\textsc{\fontsize{6}{7}\selectfont standard features}}\\\hline
		
		\textsc{\fontsize{6}{7}\selectfont kaldi clean-condition recipe} & $\checkmark$ &		 \cellcolor{orange!10} $\checkmark$ & & & $\checkmark$ & & & & &\\
		\textsc{\fontsize{6}{7}\selectfont kaldi multi-condition recipe} & & \cellcolor{orange!10} & $\checkmark$ & $\checkmark$ &  & $\checkmark$ & $\checkmark$ & $\checkmark$ & $\checkmark$ & $\checkmark$ \\
		\textsc{\fontsize{6}{7}\selectfont augmentation (algorithms~\ref{alg:bandlimited}-~\ref{alg:noisy-rir})} & & \cellcolor{orange!10} $\checkmark$ & & & & & && \\\cline{1-11} 
		\multirow{1}{*}{\textsc{test set}~\textbackslash~\textsc{neural architecture}}  & \multicolumn{3}{c|}{\textsc{\fontsize{6}{7}\selectfont parznets 2d}}  & \textsc{\fontsize{6}{7}\selectfont sincnet} & \multicolumn{2}{c|}{\textsc{\fontsize{6}{7}\selectfont cldnn}~\cite{sainath2015} } & \textsc{ \thead[c]{\fontsize{6}{7}\selectfont mfcc \\ \fontsize{6}{7}\selectfont mlp}} & \textsc{ \thead[c]{\fontsize{6}{7}\selectfont fmllr \\ \fontsize{6}{7}\selectfont mlp}} & \textsc{ \thead[c]{\fontsize{6}{7}\selectfont vdcnn \\ \fontsize{6}{7}\selectfont \cite{vdcnn}}} & \textsc{ \thead[c]{\fontsize{6}{7}\selectfont octcnn \\\fontsize{6}{7}\selectfont ~\cite{moctcnn}}}  \\
		\hline
		
		\hline
		\textsc{a: clean speech \& train mic.} & $1.96$ & \cellcolor{orange!10} $2.54$ & $2.32$ & $3.12$ & $3.17$ & $3.19$ & $4.28$ & $3.34$ & $3.27$ & $2.32$\\
		\hline
		\textsc{b: noisy speech \& train mic.}  & $18.91$ & \cellcolor{orange!10} $6.30$  & $4.38$ & $5.97$ & $33.34$ & $6.08$ & $7.44$ & $6.27$ & $5.61$ & $4.73$\\
		\hline
		\textsc{c: clean speech \& different mic.} & $19.47$ & \cellcolor{orange!10} $8.03$ & $4.30$ & $5.68$ & $16.16$ & $6.57$ & $8.73$ & $5.74$ & $5.32$ & $4.24$\\
		
		\hline
		\textsc{d: noisy speech \& different mic.} & $40.46$ & \cellcolor{orange!10} $15.88$ & $12.73$ & $16.58$ & $45.67$ & $14.06$ & $18.71$ & $16.04$ & $13.52$ & $13.57$\\
		
		\hline
		\textsc{average wer} & $26.98$ & \cellcolor{orange!10} $10.26$ & $7.80$ & $10.29$ & $35.24$ & $9.33$ & $12.14$ & $10.21$ & $8.81$ & $8.31$\\
		\hline
		
		\hline
	\end{tabular}
	\label{tbl:aurora4-multi}\vspace{-2ex}
\end{table*}

\subsubsection*{Algorithm~\ref{alg:noisy-rir} (\textsc{rir})} The sixth column in Table~\ref{tbl:aurora4} summarizes the results 
when the \textsc{rir} contribution is removed from the training sample. Overall, this augmentation scheme contributes to an approximately $10\%$ relative improvement in the WER. It is quite effective on noisy samples, independently of the microphone used. However, this comes at the expense of a slight performance degradation on clean speech that is recorded using different microphones ($4\%$ relative). This means that the augmentation scheme can scramble useful associations between speech frames and sub-phonetic units, which can be remedied via other schemes.

\subsubsection*{\textsc{gauss}} The goal of this experiment is to quantify the influence of additive Gaussian noise when coupled with more structured transformations of the inputs (i.e., Algorithms~\ref{alg:bandlimited}-~\ref{alg:noisy-rir}). The empirical results in column seven of Table~\ref{tbl:aurora4} show improvement in performance when the contribution of this transformation is \emph{removed} from the training sample. This means that further corruption with unstructured additive Gaussian noise does not help with spurious correlations. On the contrary, it degrades the performance by removing useful associations between inputs and corresponding labels.

We conclude by comparing the best model trained with the help of data augmentation (column seven in Table~\ref{tbl:aurora4}) to the one trained using the Kaldi clean-condition recipe alone (see the last column in Table~\ref{tbl:aurora4}). Our empirical evidence indicates a \emph{significant performance improvement} ($>150\%$ relative), with augmented model achieving \textsc{wer} $10.26\%$ vs. $26.98\%$ obtained using the clean-condition training.

\subsection{Potential of the Augmentation Schemes Relative to Matching of Acoustic Conditions between Training and Test Folds}
\label{subsec:multi-cond-relative}

In this section, we evaluate the potential of the proposed augmentation schemes for out-of-distribution generalization and robustness to unseen noise conditions. To this end, we exploit the aforementioned multi-condition recipe/setting for \textsc{aurora4} and compare the numbers obtained in (augmented) clean-condition training to the ones reported in~\cite{parznets20} for the former setting. Table~\ref{tbl:aurora4-multi} summarizes the results of this comparison.

The results indicate that when training with the clean-condition recipe, augmented using schemes from Section~\ref{sec:data-augmentation}, one can learn acoustic models that are competitive with those learned using the multi-condition training recipe, for which an \emph{explicit matching} (with respect to the noise types and used microphones) between train and test conditions has been performed.
Here it is important to point out that \textsc{wer} achieved by \textsc{parznets 2d} in the multi-condition setting is difficult to beat by purely acoustic models trained via the clean-condition recipe (i.e., that is one of the most competitive baselines), given the \emph{explicit matching} of training and test conditions. 

Our discussion concludes with the observation that the performance of the augmented \textsc{parznets 2d} model is also competitive with recently proposed highly effective feedforward architectures based on standard non-adaptive features, trained using a \emph{matched} sample provided by the multi-condition recipe. We leave it for future work to further tune the proposed augmentation schemes, which have been simulated with modest number of bandpass filters and hyper-parameters.

\subsection{Robustness Relative to Perturbations of the Test Folds}
\label{subsec:test-robustness}

This set of experiments is motivated by the considerations in Section~\ref{sec:theory}, where we demonstrated that perturbations of an input speech frame sampled from the vicinal density will concentrate in the feature space given by the pre-softmax layer of a neural network. Moreover, the embeddings of perturbations will concentrate in that space around the vector representing the input frame. As a result, a neural network will assign similar conditional distributions of labels given a speech frame for a neighborhood specified by the data augmentation principles. This, in particular, refers to the concentration bound from Theorem~\ref{thm: local-robust} and our derivations for inductive bias (see Section~\ref{subsec:inductive-bias}). Thus, if we perturb a test speech frame with the proposed augmentation schemes then the corresponding conditional probability vectors should be aligned in robust models. To assess this, we take the best performing model that employs the proposed augmentation schemes (column seven in Table~\ref{tbl:aurora4}) and evaluate its stability relative to perturbations of the test utterances. We also do the same experiment for the model obtained via multi-condition training. When comparing the performance of the two models, we will use the relative degradation in word error rates as a result of hypothesis smoothing over neighborhoods assigned to test utterances.

When assessing the stability, we generate $s$ perturbations of a test speech frame $x \in \mathcal{X}$ and proceed to decoding with the average conditional probability over perturbations, i.e.,
\begin{align*}
q\brackets{y \mid x} \coloneqq \frac{1}{s} \sum_{i=1}^s p\brackets{y \mid \mathcal{T}_i \brackets{x}, \alpha, W} \ ,
\end{align*}
where $\mathcal{T}$ denotes a transformation from Algorithms~\ref{alg:bandlimited}-~\ref{alg:noisy-rir}. We note here that the training fold in the Kaldi clean speech recipe for Aurora4 consists of clean speech/utterances alone. The test fold consists of noisy speech (unseen during training), as described in the first column of Table 1. In this experiment, we are interested in what happens around test points and whether the conditional probabilities of labels given frames are aligned in the vicinity of test utterances. As there is already a strong discrepancy between training and test folds, the fact that we apply Algorithms 1-4 to obtain realistic samples from the neighborhoods of tests utterances does not help our approach in any way. As evidenced with the model trained on clean speech alone (see Table~\ref{tbl:aurora4}, WER $26.98\%$), it is challenging not to fail on the original test utterances let alone their perturbations.

\begin{table}[t]
	\centering\fontsize{7}{9}\selectfont  
	\caption{The stability assessment for the \textsc{parznets 2d} models trained using: \emph{i}) the Kaldi clean-condition recipe and proposed data augmentation schemes, and \emph{ii}) the Kaldi multi-condition recipe. The evaluation is performed using the original test fold of \textsc{aurora4} and two perturbation settings realized by data augmentation. The results are summarized across groups of test sets \textsc{a-d}, as described in Table~\ref{tbl:aurora4-multi}.}
	\begin{tabular}[t]{l|c|c|c|c|c|c}
		 & \textsc{a} & \textsc{b} & \textsc{c} & \textsc{d} & \textsc{avg} & $\downarrow$~\textsc{rel}\\\hline
		\multicolumn{6}{l}{\textsc{\fontsize{7.5}{10}\selectfont parznets 2d}}\\
		\multicolumn{6}{l}{\textsc{\fontsize{7.5}{10}\selectfont kaldi clean-condition recipe \& algorithms $1$-$4$}}\\\hline
		\textsc{original} & $~2.54$ & $~6.30$ & $~8.03$ & $15.88$ & $10.26$ & $-$ \\\hline
		\textsc{smoothed (x 4)} & $~2.80$ & $~6.86$ & $~8.05$ & $17.63$ & $11.27$ & $~9.8\%$ \\\hline
		\textsc{smoothed (x 40)} & $~2.91$ & $~7.18$ & $~8.14$ & $18.14$ & $11.64$ & $13.5\%$\\\hline
		\multicolumn{6}{l}{\textsc{\fontsize{7.5}{10}\selectfont parznets 2d}}\\
		\multicolumn{6}{l}{\textsc{\fontsize{7.5}{10}\selectfont kaldi multi-condition recipe}}\\\hline
		\textsc{original} & $~2.32$ & $~4.38$ & $~4.30$ & $12.73$ & $~7.80$ & $-$ \\\hline
		\textsc{smoothed (x 4)} & $~2.62$ & $~6.15$ & $~4.86$ & $17.17$ & $10.53$ & $35.0\%$ \\\hline
		\textsc{smoothed (x 40)} & $~2.71$ & $~6.48$ & $~5.10$ & $17.66$  & $10.90$  & $39.7\%$
	\end{tabular}
	\label{tbl:aurora4-test}\vspace{-2ex}
\end{table}

Table~\ref{tbl:aurora4-test} summarizes the results of these experiments. The word error rates of the two models over original test folds (without test perturbations) are reported in rows labeled as \textsc{original}. In rows labeled as \textsc{smoothed} we report the error rates for two sets of experiments: \emph{i}) adding a single perturbation for each of the four augmentation algorithms (denoted with $\times 4$), and \emph{ii}) adding $10$ perturbations for each of the four augmentation algorithms (denoted with $\times 40$).

Our empirical results demonstrate that the acoustic model learned by training using the proposed augmentation schemes exhibits a fair amount of robustness, with only a $13.5\%$ relative performance degradation under significant perturbations (see \textsc{smoothed} 
$\times 40$ under \textsc{clean-condition} in Table~\ref{tbl:aurora4-test}) of the already challenging test fold (due to the mismatch between training and test conditions). We note that this is only a minor degradation relative to the one between models obtained via clean- and multi-condition Kaldi recipes (see Tables~\ref{tbl:aurora4} and~\ref{tbl:aurora4-multi}). 

We have repeated the same experiment for the model obtained using the multi-condition recipe/setting (\textsc{wer} $7.80\%$) and observed that there is a $39.7\%$ relative performance degradation (see \textsc{smoothed}  
$\times 40$ in Table~\ref{tbl:aurora4-test}), which indicates that the model learned using our augmentation schemes might be a better choice for unseen noise environments.

\begin{table*}[!htb]
	\centering\fontsize{7}{9}\selectfont  
	\caption{The word error rates (\%) obtained on different test sets of \textsc{aurora4} by training a \textsc{mlp} model using Kaldi clean-condition recipe with filterbank features. The columns labelled \textsc{clean} report results using the data in the original Kaldi recipe. Other columns report the results of experiments where at the onset of each epoch with probability $p=0.2$ the original utterance is retained, otherwise it is perturbed using the listed augmentations schemes.}
	\begin{tabular}[t]{l|c|c|c|c|c|c|c|c|c|c|c|c}
		& \multicolumn{3}{c|}{\textsc{\fontsize{6}{7}\selectfont kaldi clean-condition recipe}} & \multicolumn{3}{c|}{\textsc{\fontsize{6}{7}\selectfont +algorithms $1$-$4$}} & \multicolumn{3}{c|}{\textsc{\fontsize{6}{7}\selectfont +specaugment (mag. spect.)}} & \multicolumn{3}{c}{\textsc{\fontsize{6}{7}\selectfont +specaugment (fbank)}}\\\hline
		
		\textsc{\fontsize{6}{7}\selectfont pre-emphasis} & &  & $\checkmark$  &  &  & $\checkmark$   & &   & $\checkmark$ &   & & $\checkmark$ \\
		\textsc{\fontsize{6}{7}\selectfont utterance normalization}  & & $\checkmark$ & $\checkmark$  &  & $\checkmark$ & $\checkmark$   & & $\checkmark$  & $\checkmark$ &  & $\checkmark$  & $\checkmark$  \\\cline{1-13}
 
		\textsc{a: clean speech \& train mic.} & $3.68$ & $3.42$ & $3.27$ & $4.80$ & $4.17$ & $3.92$ & $5.27$ & $3.49$ & $3.72$ & $5.06$ & $4.18$ & $4.04$\\\hline
		\textsc{b: noisy speech \& train mic.} & $28.71$ & $14.78$ & $12.02$ & $16.91$ & $9.67$ & $9.31$ & $31.50$ & $14.62$ & $13.49$ & $37.53$ & $16.10$ & $15.76$\\\hline
		\textsc{c: clean speech \& different mic.} & $37.23$ & $17.00$ & $18.20$ & $21.89$ & $9.36$ & $9.77$ & $35.29$ & $17.07$ & $17.26$ & $36.60$ & $14.57$ & $15.45$\\\hline
		\textsc{d: noisy speech \& different mic.} & $54.05$ & $31.54$ & $29.01$ & $36.58$ & $20.28$ & $19.72$ & $55.22$ & $29.90$ & $28.96$ & $58.92$ & $30.51$ & $30.91$\\\hline
		\textsc{average wer} & $38.39$ & $21.39$ & $19.12$ & $24.83$ & $13.80$ & $13.42$ & $40.06$ & $20.55$ & $19.69$ & $44.31$ & $21.31$ & $21.39$\\
		\hline
		
		\hline
	\end{tabular}
	\label{tbl:aurora4-fbank}\vspace{-2ex}
\end{table*}

\subsection{Robustness Relative to \textsc{fbank} and \textsc{specaugment}}
\label{subsec:fbank}

In this section, we evaluate the effectiveness of the proposed augmentation schemes relative to the \textsc{specaugment} baseline~\cite{specaugment}. 
As our focus is on robustness relative to different types and levels of additive noise unseen during training as well as microphone effects, we again run experiments on \textsc{aurora4} using the Kaldi clean-condition recipe. \textsc{specaugment} is designed to operate on log-mel-spectrograms and in order to have an objective assessment of its effectiveness we simulate the experiment using \textsc{fbank} features. We extract in total $64$ features per frame using $25$ ms long frames and $10$ ms stride between them. This is the feature extraction setting identical to the one used by \textsc{hmm-dnn} model supplying the alignments. To simplify the experimental setting, we opt for a multi-layer perceptron (\textsc{mlp}) with five hidden layers and \textsc{relu} activation function as our \textsc{dnn} model. After linear operators in each of the hidden layers we apply batch normalization and Bernoulli dropout with $p=0.15$.

In a typical training regime with schemes such as \textsc{specaugment}, the training utterances are corrupted at the onset of each epoch. We follow this procedure and with uniform probability ($p=0.2$) decide whether to retain the original utterance or perturb it via Algorithms~\ref{alg:bandlimited}-\ref{alg:noisy-rir}. For \textsc{specaugment} we apply a similar rule and retain the original utterance approximately $20\%$ of the times. We have opted for context size of $5$ frames around a center frame and, thus, the input to \textsc{mlp} consists of $11$ successive frames. As a result of this, we simulate \textsc{specaugment} with the maximal time mask size of $10$ frames and select at most $5$ such masks per utterance. For frequency masking we have experimented with several settings and selected the maximal frequency mask size of $16$ channels. We apply one such mask per utterance. The setting described here for \textsc{specaugment} is in line with its adaptation to feedforward models~\cite[e.g., see][]{cnn-specaugment}.

There are several possible confounding factors when assessing the effectiveness of augmentation schemes in this setting. For example, \textsc{fbank} features compress information from waveform signals and this can negatively impact robustness. In addition to this, there are strategies that can mitigate some effects of additive perturbations such as utterance level mean normalization and signal pre-emphasis --- these can preclude the actual effectiveness of a data augmentation strategy. To account for the latter two factors, we perform experiments with and without them. A detailed description of the experiments is provided in Table~\ref{tbl:aurora4-fbank}. 
Our empirical results indicate a clear improvement as a result of employing the proposed augmentation schemes. More specifically, in the setting with utterance normalization and pre-emphasis the proposed approach contributes to more than $40\%$ relative improvement in \textsc{wer}. This is increased to more than $50\%$ relative if one opts not to perform utterance normalization and signal pre-emphasis. \textsc{specaugment} fails to achieve comparable  \textsc{wer} across different settings. Our hypothesis is that this is due to the fact that \textsc{specaugment} has been devised to address the \emph{memorization} effect in recurrent neural networks and, thus, it acts more as a regularizer specific to that class of models rather than providing a means of characterizing the vicinity of input/training speech signals. Our intuition is in part confirmed with the requirement to adapt the scheme for feedforward models~\cite[e.g., see][]{cnn-specaugment}.

\begin{table}[t]
	\centering\fontsize{7.2}{9}\selectfont  
	\caption{The word error rates ($\%$) obtained by training \textsc{parznets 2d} architecture using the Kaldi training fold for \textsc{ami-ihm} and then decoding the validation and test folds of \textsc{ami-ihm} and \textsc{ami-sdm} using the learned model. We did not use i-vectors in the experiments and have trained using a cross-entropy loss function. Following the original Kaldi recipe, a $3$-\textsc{gram} language model built from the \textsc{ami} and \textsc{fisher} data was adopted.}
	\begin{tabular}[t]{l|c|c|c|c}
	\multirow{2}{*}{\textsc{trained on ami-ihm}} & \multicolumn{2}{c|}{\textsc{\fontsize{6}{7}\selectfont ami-ihm}} & \multicolumn{2}{c}{\textsc{\fontsize{6}{7}\selectfont ami-sdm}}\\\cline{2-5}
	& \textsc{\fontsize{6}{7}\selectfont dev} & \textsc{\fontsize{6}{7}\selectfont eval} & \textsc{\fontsize{6}{7}\selectfont dev} & \textsc{\fontsize{6}{7}\selectfont eval}\\\hline
   \textsc{\fontsize{6}{7}\selectfont parznets 2d} & $25.1$ & $26.4$ & $76.1$ & $85.0$\\
  \textsc{\fontsize{6}{7}\selectfont parznets 2d + algs. $1$-$4$} & $26.6$ & $29.5$ & $52.2$ & $59.3$\\\hline
  \textsc{\fontsize{6}{7}\selectfont mfcc \& mlp~\cite{ami-sdm}} & - & $32.3$ & - & $76.0$\\
 \textsc{\fontsize{6}{7}\selectfont mfcc \& mlp + sdm output layer adapt.~\cite{ami-sdm}} & - & $43.7$ & - & $57.0$\\
  \textsc{\fontsize{6}{7}\selectfont mfcc \& mlp + sdm input layer adapt.~\cite{ami-sdm}} & - & $44.9$ & - & $58.1$\\
		\hline
		
		\hline
	\end{tabular}
	\label{tbl:ami-sdm}\vspace{-2ex}
\end{table}

\subsection{Robustness Relative to Conversational Speech}
\label{subsec:ami-sdm}

\textsc{ami}~\cite{renals2007recognition}~
is a conversational speech dataset with approximately $80$ hours of speech. It comes in three parts, two of which are of interest to this work: \emph{i}) \textsc{ami-ihm} that contains data recorded using individual headset microphones, and \emph{ii}) \textsc{ami-sdm} that contains data recorded using a distant microphone. Our goal in this section is to demonstrate that the proposed augmentation schemes can be used to improve the performance of acoustic models learned on data from headset microphones in distant-talking speech recognition tasks. To this end, we first train the baseline \textsc{parznets 2d} architecture using the training fold of Kaldi \textsc{ami-ihm} recipe and evaluate its effectiveness on \textsc{ami-sdm} with distant-talking speech. This is then repeated by performing \emph{online} augmentation of training utterances using Algorithms~\ref{alg:bandlimited}-\ref{alg:noisy-rir}, i.e., at the onset of each epoch one decides with the uniform probability whether to keep the original \textsc{ami-ihm} training utterance or corrupt it via the proposed augmentation schemes. The alignments for this tasks were generated using the Kaldi \textsc{ami-ihm} recipe configured with $3,984$ \textsc{hmm} state ids. The training process runs for $16$ epochs with the remaining setup being the same as in other experiments. 

Table~\ref{tbl:ami-sdm} reports the results of this experiment. We observe that on both \textsc{ami-sdm} test folds (labelled with \textsc{dev} and \textsc{eval}) the addition of vicinal characterizations captured by Algorithms~\ref{alg:bandlimited}-\ref{alg:noisy-rir} 
improves the word error rate in excess of $43\%$ relative. Our empirical result, thus, demonstrates that the proposed schemes can improve robustness of acoustic models relative to distant-talking speech. Moreover, we compare our results to prior work~\cite{ami-sdm} 
where the goal was to improve robustness relative to distant-talking speech by means of layer adaptation using \textsc{ami-sdm} data. The empirical results show that our vicinal model is competitive with layer adaptation baselines despite relying on \textsc{ami-ihm} training utterances alone.

\section{Discussion}
\label{sec:discussion}

The main focus of prior studies on leveraging data augmentation for improving the performance of acoustic models is on achieving gains across different benchmarks datasets. This is typically done without considering the underlying machine learning principles that are responsible for translating the additional information originating from the augmentation schemes into the robustness. This work aims to bridge this gap by posing data augmentation as an instance of vicinal risk minization. The latter is a theoretically well founded setting that allows for an insight into the inductive bias incorporated into neural networks by means for augmentation schemes. More specifically, we have demonstrated in Section~\ref{sec:theory}
that for robustness relative to distribution shifts between train and test samples one requires a good characterization of the marginal density around training samples. Theorem~\ref{thm: local-robust} 
is our main theoretical contribution and it shows that the pre-softmax layer outputs that correspond to signals sampled from the vicinal density centered at a training sample concentrate in the feature space given by that layer of the neural architecture. As a result of this, neural networks will perform smoothing rather than interpolation and (in hybrid \textsc{hmm-dnn} models) assign similar likelihoods for \textsc{hmm} state ids across neighborhoods described by vicinal densities. We have further supported this via insights in Section~\ref{subsec:inductive-bias}
where it was demonstrated that learning with cross-entropy loss in the vicinal setting (i.e., with data augmentation) amounts to maximizing the log-likelihood over neighborhoods rather than individual samples. These theoretical findings were put to the test in Section~\ref{subsec:test-robustness}
where we trained a model using the Kaldi clean-condition recipe for \textsc{aurora4} and decoded the noisy/divergent test folds by employing average likelihoods relative to vicinal densities at test samples rather than standard pointwise predictions. The empirical results indicate that there is only a minor performance degradation which supports our results on inductive bias and concentration of vicinal samples. Moreover, we have evaluated in the same manner a model trained using the Kaldi multi-condition recipe for \textsc{aurora4} and observed a significant performance degradation relative to samples from vicinal densities (see Table~\ref{tbl:aurora4-test}). 

In the second part of the paper, we focus on devising augmentation schemes that can be effective for learning models robust relative to different types and levels of stationary additive noise, as well as divergences due to microphone effects. This is followed by a focused empirical study that first characterizes the effectiveness of individual schemes relative to different sources of additive noise and microphone effects (Section~\ref{subsec:impact}). Our empirical results in this regard are compelling showing significant improvement in out-of-distribution generalization compared to training using the standard risk minimization principle. More specifically, we observe an improvement in excess of $150\%$ relative on \textsc{aurora4} when generalizing from clean speech to noisy. This is further strengthened with empirical results on conversational speech where we show that the proposed augmentation schemes can help in the extremely difficult problem of generalizing from data collected via headset microphones to distant-talking speech. We observe an improvement in excess of $40\%$ relative on this task (Section~\ref{subsec:ami-sdm}). We have also performed a detailed analysis of the approach relative to state-of-the-art augmentation schemes based on waveform signals and filterbank features. More specifically, the multi-condition recipe for \textsc{aurora4} offers the most effective waveform-based augmentation scheme known to us because it performs explicit matching between train and test conditions relative to types of additive noise and microphone effects. We have demonstrated that the model learned via vicinal risk minimization and the Kaldi clean-condition recipe is competitive with multi-condition training (Section~\ref{subsec:multi-cond-relative}), which is a difficult task to achieve. Moreover, we also show that the proposed approach offers means for learning effective models based on filterbank features (Section~\ref{subsec:fbank}). In that experiment, we have contrasted our augmentation schemes to \textsc{specaugment}~\cite{specaugment}
and demonstrated that we clearly outperform this baseline on the task where there is a significant divergence in acoustic properties characteristic of train and test folds. Moreover, our empirical results have identified a shortcoming of the \textsc{specaugment} scheme, which we hypothesize is due to the fact that it was designed to address the hidden state memorization in recurrent neural networks. Namely, time and frequency masking forces recurrent neural networks to infer the missing information from the available context (i.e., sequence of preceding and succeeding non-masked frames). This is probably the reason why it has been coupled mainly with recurrent architectures in prior work. Thus, the proposed augmentation schemes are not only more effective but also more general and widely applicable for achieving robustness relative to additive noise and microphone effects.

\section{Conclusion}
\label{sec:conclusion}

We have proposed an effective approach for learning robust acoustic models and demonstrated empirically that it can generalize to unseen acoustic conditions. Our theoretical contributions show that the approach can incorporate a robust inductive bias into the learning process and that it provides a flexible method for characterizing vicinal risk estimates around training observations. The latter allows for effective out-of-distribution generalization and motivates further research in the direction of vicinal risk minimization. We have also given a bound that characterizes the robustness of waveform-based models, given in terms of the Jacobian and Hessian tensors. An interesting aspect of this bound is that it can be used as a basis for regularization mechanisms, which we aim to explore further in future work. In addition to all of this, we have also proposed highly effective data augmentation schemes and demonstrated empirically that they have the potential to address the issues with spurious correlations in acoustic models. Our ablation study carefully dissects the influence of individual schemes on the out-of-distribution generalization relative to several different noise types and microphone effects.

\section*{Acknowledgments}
The authors would like to thank Erfan Loweimi for providing the results of his experiments with the \textsc{cldnn} baseline on \textsc{aurora4} during the revision period.

\appendix

\begin{proof}
	We start by expanding  
	the sufficient statistic using the Taylor theorem for multivariate functions, i.e.,
	\begin{align*}
	\begin{aligned}
	& \Psi\brackets{x + \epsilon} =&& \Psi\brackets{x} + \nabla \Psi\brackets{x}^{\top}\epsilon \ + &\\
	& &&\frac{1}{2}\sum_{i,k=1}^d\nabla^2\Psi_{ik,*}\brackets{x} \epsilon_i \epsilon_k + o(\norm{\epsilon}^2) \ . &
	\end{aligned}
	\end{align*}
	From our assumptions, it follows that there exists a constant $A>0$ such that for all $x \in \mathcal{X}$ it holds $\absolute{\nabla^2\Psi_{ik,j}\brackets{x}}<A$, with $1\leq i,k \leq d$ and $1 \leq j \leq D$. This then implies that we can upper bound the perturbation as
	\begin{align*}
	\begin{aligned}
	& \norm{\Psi\brackets{x + \epsilon} - \Psi \brackets{x}} \leq &\\ 
	& \norm{\nabla\Psi\brackets{x}^{\top}\epsilon} + \frac{1}{2}
	\norm{\sum_{i,k=1}^d\nabla^2\Psi_{ik,*}\brackets{x} \epsilon_i \epsilon_k}
	+ o(\norm{\epsilon}^2)\ . &
	\end{aligned}
	\end{align*}
	As the random variable  $\epsilon$ has law $\epsilon \sim \mathcal{N}\brackets{0, \sigma^2\mathbb{I}}$, it follows that we can bound the third term in this inequality by $o\brackets{\sigma^2}$. That term also determines the approximation order for our bound and it is independent of the neighborhood centered at $x \in \mathcal{X}$. We now focus on the two leading terms and introduce 
	a univariate random variable $Z=\norm{\nabla\Psi\brackets{x}^{\top}\epsilon} + \frac{1}{2}\norm{\sum_{i,k=1}^d\nabla^2\Psi_{ik, *}\brackets{x} \epsilon_i \epsilon_k}
	$. Then,
	\begin{align*}
	P_{\epsilon} \brackets{\norm{\Psi\brackets{x + \epsilon} - \Psi \brackets{x}} \geq r} \leq P_{\epsilon} \brackets{Z > r} < \frac{\mathbb{E}_{\epsilon}\sbrackets{Z}}{r} ,
	\end{align*}
	where the last inequality follows from the Markov bound.
	
	Now, observe that $\nabla\Psi\brackets{x}^{\top}\epsilon$ is a Gaussian random variable that follows the distribution $\mathcal{N}\brackets{0, \sigma^2\nabla\Psi\brackets{x}^{\top}\nabla\Psi\brackets{x}}$. Denote $C=\nabla\Psi\brackets{x}^{\top}\nabla\Psi\brackets{x}$ and observe that this matrix is positive semi-definite. 
	Thus, $C$ admits an eigendecomposition $C=U\Lambda U^{\top}$ with an orthogonal matrix $U$ and a diagonal eigenvalue matrix $\Lambda$. This allows us to rewrite the first term from the right-hand side of the Markov bound as
	\begin{align*}
	\begin{aligned}
	& \mathbb{E}_{\epsilon} \sbrackets{\norm{\nabla\Psi\brackets{x}^{\top}\epsilon}}= \mathbb{E}_{v \sim \mathcal{N}\brackets{0, \sigma^2\Lambda}}\sbrackets{\norm{v}} \ . &
	\end{aligned}
	\end{align*} 
	On the other hand, from the Jensen inequality it follows that
	\begin{align*}
	\brackets{\mathbb{E}_{v}\sbrackets{\norm{v}}}^2 \leq \mathbb{E}_v\sbrackets{\norm{v}^2} = \sigma^2 \tr \brackets{\Lambda} = \sigma^2\tr \brackets{C}=\sigma^2a \ ,
	\end{align*} 
	with the constant $a>0$ introduced in Definition~\ref{def:trace-constants}.
	
	For the second term, observe that~\cite{Lang}
	\begin{align*}
	\mathbb{E}_{\epsilon} \sbrackets{\norm{\sum_{i,k=1}^d\nabla^2\Psi_{ik,*}\brackets{x} \epsilon_i \epsilon_k}} \leq \mu_1\mathbb{E}_{\epsilon} \sbrackets{\norm{\epsilon}^2}  \ ,
	\end{align*}
	where $\mu_1$ is the largest singular value of the 
	$D$ tensors $\nabla^2\Psi_{**, j}\brackets{x}$ with $1\leq j \leq D$. As $\epsilon$ is an isotropic Gaussian random variable, we have 
	$\mathbb{E}_{\epsilon} \sbrackets{\norm{\epsilon}^2}=\sigma^2 \cdot \mathbb{E}_{u\sim \mathcal{N}\brackets{0,\mathbb{I}_d}} \sbrackets{\norm{u}^2}=\sigma^2d $.
	While this provides a bound on the second term, there is an undesirable dependence on the dimension of the instance space $d$. In the following part of the proof, we show how this can be avoided by using the trace of the Hessian tensor as captured by the constant $b$ introduced in Definition~\ref{def:trace-constants}. 
	
	First, we observe that
	\begin{align*}
	\begin{aligned}
	& \mathbb{E}_{\epsilon} \sbrackets{\norm{\sum_{i,k=1}^d \nabla^2 \Psi_{ik, *}\brackets{x} \epsilon_i \epsilon_k}^2 } = & \\
	& \sum_{j=1}^D \mathbb{E}_{\epsilon} \sbrackets{ \brackets{\epsilon^{\top} \nabla^2\Psi_{**, j}\brackets{x} \epsilon}^2 } = \sum_{j=1}^D \mathbb{E}_{v \sim \mathcal{N}\sbrackets{0, \sigma^2\mathbb{I}}} \brackets{v^{\top}\Xi_j v}^2 = & \\
	& \sum_{j=1}^D \mathbb{E}_{v \sim \mathcal{N}\brackets{0, \sigma^2\mathbb{I}}} \sbrackets{\brackets{\sum_{k=1}^d \xi_{jk} v_{k}^2}^2 }= & \\
	& 2\sigma^4 \sum_{j=1}^D \tr \brackets{\nabla^2\Psi_{**, j}\brackets{x}\nabla^2\Psi_{**, j}\brackets{x}} + \tr\brackets{\nabla^2\Psi_{**, j}\brackets{x}}^2 \ ,&
	\end{aligned}
	\end{align*}
	where $\Xi_j$ is a diagonal eigenvalue matrix with entries $\xi_{jk}$ and $1 \leq k \leq d$. The latter eigenvalues are from the decomposition of the symmetric matrix $\nabla^2 \Psi_{**, j}\brackets{x} \in \mathbb{R}^{d\times d}$.
	Now, combining the latter with the Jensen inequality we have the following upper bound on the second order term
	\begin{align*}
	\mathbb{E}_{\epsilon} \sbrackets{\norm{\sum_{i,k=1}^d\nabla^2\Psi_{ik,*}\brackets{x} \epsilon_i \epsilon_k}} \leq \sigma^2 \sqrt{2b} \ .
	\end{align*}
	Hence, we have that it holds
	\begin{align}
	\mathbb{E}_{\epsilon}\sbrackets{Z} < \sigma\brackets{\sqrt{a}+\sigma\sqrt{\nicefrac{b}{2}}} \ .
	\end{align}
	This then implies that for all $\delta > 0$ and $r > \frac{\sigma}{\delta}\brackets{\sqrt{a} + \sigma\sqrt{\nicefrac{b}{2}}}$
	\begin{align*}
	P_{\epsilon}\brackets{\norm{\Psi\brackets{x + \epsilon} - \Psi \brackets{x}} \geq r} < \delta \ .
	\end{align*}
\end{proof}

\noindent {\bf Remark.} The result in Theorem~\ref{thm: local-robust} applies within the second order Taylor expansion, due to the extra term that was omitted from consideration and which can be bounded by $o\brackets{\sigma^2}$.

{
	\small
	\bibliographystyle{IEEEtran}
	\bibliography{parznets}
}

\end{document}